\documentclass[aps,prb,twocolumn,showpacs,superscriptaddress]{revtex4}
 \pdfoutput=1
\usepackage{graphicx}
\usepackage{epsf} 
\usepackage{dcolumn}
\usepackage{bm}
\usepackage{amssymb}
\usepackage{amsmath}
\usepackage{amsbsy}
\usepackage{graphicx,color,ifthen}
\usepackage[latin1]{inputenc}
\usepackage{epstopdf}

\usepackage{subfig}

\begin{document}

\title[]{Symmetry of surface nanopatterns induced by ion-beam sputtering: 
the role of anisotropic surface diffusion}
\author{Javier Renedo}
\affiliation{Instituto de Investigaci\'{o}n Tecnol\'{o}gica (IIT), Universidad Pontificia Comillas, 28015 Madrid, Spain}
\author{Javier Mu\~noz-Garc\'ia}
\affiliation{Departamento de Matem\'aticas and Grupo Interdisciplinar de Sistemas Complejos (GISC), Universidad Carlos III de Madrid, 28911 Legan\'es, Spain}

\author{Mario Castro}
\address{GISC and Grupo de Din\'amica No Lineal (DNL), Escuela T\'ecnica Superior de Ingenier{\'\i}a (ICAI), \\
Universidad Pontificia Comillas, 28015 Madrid, Spain}

\author{Rodolfo Cuerno}
\affiliation{Departamento de Matem\'aticas and Grupo Interdisciplinar de Sistemas Complejos (GISC), Universidad Carlos III de Madrid, 28911 Legan\'es, Spain}


\date{\today}

\begin{abstract}
Ion Beam Sputtering (IBS) is a cost-effective technique able to produce ordered nanopatterns on the surfaces of different materials.
To date, most theoretical studies of this process have focused on systems which become amorphous under irradiation, e.g.\ semiconductors at room temperature. Thus, in spite of the large amount of experimental work on metals, or more recently on semiconductors at high temperatures, such experimental contexts have received relatively little theoretical attention. These systems are characterized by transport mechanisms, e.g.\ surface diffusion, which are anisotropic as a reflection of the crystalline structure not being overruled by the irradiation. Here, we generalize a previous continuum theory of IBS at normal incidence, in order to account for anisotropic surface diffusion. We explore systematically our generalized model in order to understand the role of anisotropy in the space ordering properties of the resulting patterns. In particular, we derive a height equation which predicts morphological transitions among hexagonal and rectangular patterns as a function of system parameters and employ an angular correlation function to assess these pattern symmetries. By suitably choosing experimental conditions, it is found that one might be able to experimentally control the type of order displayed by the patterns produced.
\end{abstract}


\maketitle

\section{Introduction} \label{Introduction}

%
%
%
%

Ion-beam sputtering (IBS) is a technique employed to efficiently nanostructure surfaces: \cite{Som2013} a solid target is bombarded with energetic ions, which erode material inducing self-organized pattern formation at the target surface. There is a wide technological interest in this technique, since it allows to obtain ordered nanostructures with controlled roughness, wavelength, and orientation. \cite{BuatierdeMongeot2009,Munoz-Garcia2009book} Moreover, it is scalable, cost-efficient, and can be used in many materials, including semiconductors, metals, and insulators. One challenge that still limits the widespread use of IBS is the lack of an unified theoretical framework which guides experimental designs.

In this regard, continuum models have been relatively successful in describing the dynamical behavior of these nanostructures, typically in terms of macroscopic variables like the target surface height. For materials which are or become amorphous under low energy ($E \simeq 1$ keV) IBS, like semiconductors, \cite{Gnaser1999} Bradley and Harper (BH) pioneered this approach through a linear continuum theory which explains the formation of ripples and their orientation,\cite{Bradley1988} based on Sigmund's theory of sputtering \cite{Sigmund1969} and Mullins' thermal surface diffusion.\cite{Mullins1957} The success of this model to account for the origin of the patterns triggered an intense activity and further generalizations. In particular, relevant nonlinear corrections were identified in Ref.~\onlinecite{Cuerno1995}, leading to an equation of the Kuramoto-Sivashinsky (KS) type. \cite{Cuerno2011} Importantly, nonlinearities were seen to moderate the pattern-forming linear instability and eventually stabilize the surface morphology.

The BH equation and its generalizations were similarly derived as in Ref.\ \onlinecite{Bradley1988}, by adding together physically-diverse contributions into a single equation for the target height. Alternatively, as shown in Refs.~\onlinecite{Aste2004} and \onlinecite{Aste2005}, one can describe the dynamics of two different fields, the surface height and the density of material (e.g.\ adatoms, advacancies) subject to transport at the surface. This approach describes surface dynamics successfully in many different contexts, from granular matter \cite{Csahok2000} to epitaxial growth.\cite{Tiedje2008} In the IBS context, it enables improvements,\cite{Cuerno2011} most notably by coupling different physical mechanisms in a natural way. For instance irradiation is expected to influence surface diffusion and be reflected in the corresponding terms in the height equation, typically as a linear high-order derivative term. However, direct expansion of Sigmund's contribution in the erosion velocity to such linear \cite{Makeev1997,Makeev2002} or nonlinear orders \cite{Kim2004,Kim2005} are affected by consistency issues with respect to pattern formation.\cite{Castro2005b,More2012,Bradley2011a} Such type of issues do not occur in two-field formulations.\cite{Cuerno2011} Thus, the KS equation was consistently generalized into the so-called extended KS (eKS) model for IBS.\cite{Castro2005,Munoz-Garcia2006} For normal incidence conditions, this model has been studied for one-dimensional (1D) systems,\cite{Munoz-Garcia2006b}  and for 2D systems and rotating targets.\cite{Munoz-Garcia2009} Oblique incidence is studied in Ref.\ \onlinecite{Munoz-Garcia2008}. While being a phenomenological approximation of fuller hydrodynamic descriptions, \cite{Castro2012a} two-field modeling provides a generic framework which allows to modify the interface equation when improved models of erosion and/or transport are considered. To date, the two-field model and/or the eKS equation have been (semi)quantitatively validated in several IBS experiments.\cite{Gago2006,Munoz-Garcia2010,Bikondoa2012,Kim2013,Kramczynski2014}

The scenario just described focuses almost exclusively on targets for which the crystalline structure is overruled by the IBS process. However, there are important instances in which this
is not the case, most notably metals \cite{Valbusa2002,BuatierdeMongeot2009} and semiconductors at high temperature.\cite{Ou2013,Ou2014,Ou2015} In both cases, the strong dependence of the diffusivities of ad-atoms and vacancies with the crystallographic direction can play a crucial role in the pattern formation process. For metals, the surface is not amorphized after ion impact.
For semiconductors, increasing temperatures above the recrystallization value analogously restores dynamical dominance of crystalline anisotropies.
As discussed in Refs.~\onlinecite{Valbusa2002} and \onlinecite{Chan2007} for metallic systems, two regimes can be distinguished: (i) \emph{diffusive regime}, when pattern formation is governed by thermal surface diffusion, typically for intermediate temperatures and relatively low ion fluxes, and (ii) \emph{erosive regime}, when pattern formation is controlled by the direction of the ion beam, usually for very high or very low temperatures and for large enough ion fluxes. For instance, the diffusive regime allows for anisotropic ripple formation under isotropic, normal incidence conditions, and in general implies that both the ripple wavelength and orientation are controlled by temperature. \cite{Valbusa2002} This behavior can not be explained using previous models of IBS for amorphous targets, in which the diffusive terms are isotropic. A generalization of the linear BH model to anisotropic materials was proposed in Refs.~\onlinecite{Rusponi1998a} and \onlinecite{Costantini2001b}. Some properties observed in IBS of metals could thus be described, but in this formulation surface transport does not couple with erosion in a natural way.
Likewise, with a focus on strongly kinetic effects, previous two-field \cite{Aste2004,Aste2005} and one-field models \cite{Ou2013,Ou2014,Ou2015} have described crystalline anisotropies, but only at nonlinear order. However, in principle under these conditions surface diffusion currents need to include anisotropic linear terms, \cite{Golubovic2002,Levandovsky2006} which account for e.g.\ the direction-dependence of barriers to adatom/advacancy diffusion on terraces, along step edges, etc.\cite{Misbah2010}

In view of the above, there is a need for studies in which crystalline anisotropies to material transport are systematically addressed, for surfaces undergoing low energy IBS. Already the simplest scenario of anisotropic linear surface diffusion can lead to non-trivial modifications of pattern properties, even if possibly not modifying other, such as stability phase diagrams.\cite{Golubovic2002,Levandovsky2006} For instance, recent experiments with gold targets \cite{Kim2011,Kim2013} have obtained highly ordered nanodot patterns by sequential ion-beam sputtering (SIBS). The procedure consisted in sputtering under normal incidence a pre-patterned ripple structure previously obtained by oblique bombardment. When the initial surface is flat and not pre-patterned, a more disordered dot pattern is obtained, which still shows square in-plane order.\cite{Kim2011} 
Although the (isotropic) eKS model reproduces many of the experimental properties of the ensuing nanobead pattern,\cite{Kim2011,Kim2013} it is not able to predict this square symmetry, being limited to describing more isotropic, hexagonal order.


In this paper we put forward a two-field model of IBS nanopatterning under conditions in which anisotropies to surface transport are relevant. As a basis for further studies, our goal is to demonstrate non-trivial effects arising already within the simplest anisotropic scenarios, which will motivate our choices in the modeling of both, transport and irradiation-related mechanisms. As a result, we obtain a generalization of the eKS equation, which is integrated numerically for normal ion incidence. Our results show that anisotropic surface diffusion has non-trivial effects, and allows to reproduce nanopatterns with different local ordering structures in monoelemental systems, from hexagonal to square, akin to those experimentally reported for IBS of metals.\cite{Kim2011,Kim2013}

This paper is organized as follows. Our generalized two-field model with anisotropic diffusion is put forward in Section \ref{Model}. In principle, the model holds for arbitrarily oblique ion incidence. However, in order to isolate the effect of anisotropy in diffusion, rather than in irradiation, we then restrict ourselves to normal incidence. For this case we derive an equivalent interface equation which generalizes the eKS model. This novel nonlinear equation is studied numerically in Section \ref{sec:simulation}, where the effect of each one of the parameters which control the system behavior is discussed in detail. Finally, Section \ref{Concl} contains our conclusions and an outlook on future developments. Some details on our modeling are provided in the Appendix.


\section{Generalized Two-field Model} \label{Model}
\subsection{Derivation}

A two-field model is a system of two coupled partial differential equations describing the temporal evolution of two important macroscopic variables.\cite{Cuerno2011} The first variable corresponds to the height of the bombarded surface, $h (\boldsymbol{x}, t)$, at substrate position $\boldsymbol{x} = (x,y)$ and time $t$. The second one describes the thickness (which, for a fixed atomic volume, is proportional to the density) of the mobile surface adatoms layer, $R (\boldsymbol{x}, t)$. For semiconductors at room temperature, irradiation creates an amorphous layer with a thickness of the order of the ion range,\cite{Gnaser1999} within which transport can be described in terms of viscous flow.\cite{Castro2012a,Norris2012,Castro2012b} However, for metals or for semiconductors at high temperatures, such amorphization does not take place, \cite{Valbusa2002,Chan2007} so that the surface layer on which transport occurs can be assumed to have roughly an atomic thickness. In such cases, the dynamics of $h$ and $R$ are coupled by mass conservation as
\begin{align}
	\label{eq:couR}
	\frac{\partial R}{\partial t} &= (1 - \phi) \Gamma_{ex} - \Gamma_{ad} - \nabla \cdot \boldsymbol{J}, \\
	\label{eq:couH}
	\frac{\partial h}{\partial t} &= - \Gamma_{ex} + \Gamma_{ad}.
\end{align}
Here, $\Gamma_{ex}$ is a function that describes the rate at which target atoms are excavated (locally decreasing $h$) and can become mobile (locally increasing $R$), while $\Gamma_{ad}$ models the rate of atom addition back to the solid (increasing $h$ and decreasing $R$). The parameter $\phi \in [0, 1]$ measures the fraction of eroded atoms that are actually sputtered away from the surface, while $\bar{\phi} = 1 - \phi$ measures the fraction of eroded atoms that remain subject to transport at the surface.

Equation \eqref{eq:couR} includes an additional conserved current, $\boldsymbol{J}$, which accounts for surface transport mechanisms. For instance, this current could readily incorporate Carter-Vishnyakov (CV) contributions \cite{Carter1996} due to mass redistribution, believed to be relevant in the case of semiconductors at room temperature.\cite{Munoz-Garcia2014} These have been employed in a number of similar two-field models for IBS of compound systems, or for IBS of monoelemental targets under concurrent impurity codeposition, see e.g.\ Ref.\ \onlinecite{Bradley2012b} for a partial overview. For monoelemental targets, CV-type effects can also be reflected in $\Gamma_{ex}$.\cite{Castro2005} In any case, for metals or semiconductors at high temperatures this type of mass redistribution is not expected to play a role, nor is, on a more mesoscopic level, surface-confined viscous flow.\cite{Umbach2001,Castro2012a,Norris2012,Castro2012b} The main transport mechanism is expected to be, rather, thermal surface diffusion. Microscopically, this is an activated process for which energetic barriers exist, whose depths depend on the crystallographic directions.\cite{Ala-Nissila2002} On a more coarse-grained level, as e.g.\ in Mullins' classic theory,\cite{Mullins1957} surface diffusion is mediated by surface tension, which for metals is paradigmatically anisotropic.\cite{Davis2001} Mathematically, we thus consider Fickian diffusion at the surface as described by
\begin{equation}
\boldsymbol{J} = - \boldsymbol{D} \nabla R, \label{eq:J}
\end{equation}
where $\boldsymbol{D} \in \mathbb{R}^{2 \times 2}$ is a (positive definite) diffusion tensor, rather than a constant, that implements the present type of anisotropy. Its most general form reads
\begin{equation}\label{eq:dif_tensorD_gral}
	\boldsymbol{D} =
	\boldsymbol{M}(\psi) \left[
	\begin{array}{cc}
		D_{\|} & 0 \\
		0 & D_{\bot}
	\end{array}
	\right] \boldsymbol{M}^{-1} (\psi)=
    \left[ \begin{array}{cc}
		D_{xx} & D_{xy} \\
		D_{xy} & D_{yy}
	\end{array}
	\right],
\end{equation}
where $\boldsymbol{M}(\psi)$ is a counterclockwise rotation matrix of angle $\psi$ which gives the orientation of the fast diffusion direction with respect to the $\boldsymbol{\hat{x}}$ direction, so that $D_{\|} \geq D_{\bot} >0$ without loss of generality. 

Similar models of anisotropic surface diffusion have been employed e.g.\ in models of the dynamics of vicinal \cite{Danker2004} and singular \cite{Meca2013} surfaces in epitaxy, which have been experimentally validated.\cite{Meca2013b} Note that the present (surface diffusion) anisotropy is independent of that induced by ion bombardment under an oblique angle of incidence; consequently, it is still relevant under otherwise isotropic normal incidence conditions. In principle, $\boldsymbol{J}$ could incorporate additional terms, most notably (linear and nonlinear)  contributions depending on the surface height, which are related with Ehrlich-Schwoebel (ES) anisotropic barriers to surface diffusion.\cite{Michely2004,Misbah2010,Ou2013,Ou2014,Ou2015} However, the morphological instability associated with these terms differs physically from the BH instability. In order to assess more clearly the interplay between anisotropic surface diffusion and IBS, at this stage ES-related mechanisms are left for further work. They are expected to play a significant role in patterns whose in-plane order extends to a longer range, and in which wavelength coarsening is more sizeable, than in e.g.\ experiments on nanobead formation.\cite{Kim2011,Kim2013}


In order to close the system of equations \eqref{eq:couR}-\eqref{eq:couH}, the excavation and addition rates have to be related to the density of adatoms ($R$) and to the geometry of the substrate ($h$ and its space derivatives) themselves. For an arbitrary incidence angle $\theta$ and assuming that the projection of the ion beam is along the $\boldsymbol{\hat{x}}$ direction, we consider\cite{Castro2005,Munoz-Garcia2006}
\begin{eqnarray}\label{eq:Gamma_ex}
\Gamma_{ex} = \alpha_0 \Big[ 1 + \alpha_{1x} \frac{\partial h}{\partial x} + \alpha_{2x}\frac{\partial^2 h}{\partial x^2} + \alpha_{2y} \frac{\partial^2 h}{\partial y^2} \\ \nonumber + \alpha_{3 x} \Big( \frac{\partial h}{\partial x} \Big)^2  +  \alpha_{3y} \Big( \frac{\partial h}{\partial y} \Big)^2 \Big] \mbox{\space ,} 
\end{eqnarray}
where $\alpha_0>0$ is the excavation rate for a flat surface. In Eq.\ \eqref{eq:Gamma_ex} the terms with coefficients $\alpha_{1x}$, $\alpha_{2j}$ correspond to the lowest linear-order approximation to the dependence of the sputtering yield on the local height derivatives, as in BH's theory,\cite{Bradley1988} while those with coefficients $\alpha_{3j}$ characterize the corresponding lowest-order nonlinear corrections.\cite{Cuerno1995} Due to the assumed geometry for ion bombardment, for normal incidence ($\theta=0$) one has $\alpha_{1x}=0$, while $\alpha_{2x}=\alpha_{2y}$, $\alpha_{3x}=\alpha_{3y}$. In general, in the absence of CV-type effects, one expects $\alpha_{2j}<0$ leading to pattern formation (BH instability), while non-zero $\alpha_{3j}$ guarantee non-exponential increase of the surface roughness for long times, as mentioned above.

Finally, for the local addition rate we consider\cite{Munoz-Garcia2006}
\begin{equation}\label{eq:Gamma_ad}
	\Gamma_{ad} = \gamma_0 \Big[  R \Big( 1 + \gamma_{2x} \frac{\partial^2 h}{\partial x^2} + \gamma_{2y} \frac{\partial^2 h}{\partial y^2} \Big) - R_{eq} \Big]  \mbox{\space ,}
\end{equation}
where $\gamma_0>0$ is the nucleation rate, i.e., $1/\gamma_0$ represents the average time in which ad-atoms incorporate to a flat surface. As discussed earlier,\cite{Munoz-Garcia2008,Munoz-Garcia2009} in absence of ion-beam driving Eq.\ \eqref{eq:Gamma_ad} describes Mullins' thermal surface diffusion, in such a way that $R_{eq}$ is related with the surface concentration of mobile species, while $\gamma_{2j}\ge 0$ are surface tension coefficients which, in general, can also be anisotropic.


The two-field model \eqref{eq:couR}-\eqref{eq:Gamma_ad} supports a flat solution in which the surface height erodes with an uniform speed and both $h$ and $R$ are space-independent functions.\cite{Renedo2013} Performing a standard linear stability analysis of perturbations of this solution which are periodic with wave-vector $\boldsymbol{k}$, we can obtain the pattern wavelength, $\ell_i$, along each direction $i =x, y$ within linear approximation. Specifically, we define $\ell_i = {2 \pi}/{ k_i^{\ell}}$ as the length-scales at which the dispersion relation is maximized in each direction. In our case,\cite{Renedo2013} $k_{i}^\ell = \left({\epsilon \phi \gamma_0 \alpha_{2 i}}/{2 R_{eq} D_{i} \gamma_{2i}}\right)^{1/2}$,
where the parameter $\epsilon \equiv \alpha_0/\gamma_0$ turns out to be small as a consequence of the difference between the typical time scales associated with diffusion (typically of the order of ps) and the ion-beam driving (of the order of seconds for the ion fluxes usually employed).\cite{Munoz-Garcia2008} This separation in time scales allows one to simplify the analysis of the mathematical model [Eqs.\ \eqref{eq:couR}-\eqref{eq:couH}] since it allows to perform a multiple-scale perturbative analysis to obtain a closed equation for the height. Analysis that closely follows Ref.\ \onlinecite{Munoz-Garcia2008} leads to an effective nonlinear equation for the time evolution of $h$, which reads
\begin{eqnarray}\label{eq:eKS_anis}
\nonumber
\frac{\partial h}{\partial t} &=&   \gamma_x \frac{\partial h}{\partial x} +
\sum_{i=x,y} \Omega_{ij} \frac{\partial^2}{\partial i\partial j} \Big ( \frac{\partial h}{\partial x} \Big) \\
 & + & \sum_{i=x,y} \left[ -\nu_i \frac{\partial^2 h}{\partial i^2}  + \lambda_i^{(1)} \Big ( \frac{\partial h}{\partial i} \Big)^2 \right] \\ 
 & - & \sum_{i,j,k=x,y} \left[ \mathcal{K}_{ijk} \frac{\partial^2}{\partial i\partial j} \Big ( \frac{\partial^2 h}{\partial k^2} \Big ) + \lambda_{ij}^{(2)}\frac{\partial^2}{\partial i\partial j} \Big ( \frac{\partial h}{\partial k} \Big )^2 \right] , \nonumber
\end{eqnarray}
where the coefficients $\gamma_x$, $\Omega_{ij}$, $\nu_i$, $\lambda_i^{(j)}$, and $\mathcal{K}_{ijk}$ depend on ion energy, flux, incidence angle, etc.\ through their dependencies on $\alpha_{ij}$ and all other parameters entering $\Gamma_{ex}$ and $\Gamma_{ad}$, as specified in Appendix \ref{app.B}.

Equation (\ref{eq:eKS_anis}) is partially similar to the evolution equation obtained in Ref.~\onlinecite{Munoz-Garcia2006} for isotropic surface diffusion and oblique ion incidence. However, in that case the only geometrical condition responsible for breaking the $x \leftrightarrow y$ symmetry was the non-zero value of the incidence angle, in such a way that the system was symmetric under space reflection in the $y$ direction, but not in the $x$ direction. In the case of Eq.\ \eqref{eq:eKS_anis}, this same cause for space anisotropy is enhanced by anisotropic surface diffusion and by anisotropic surface tension. As a consequence, not only are the $x \leftrightarrow y$ and $x \leftrightarrow -x$ symmetries broken, but the $y\leftrightarrow -y$ symmetry is broken as well, now by the two latter conditions. The differences between Eq.\ (\ref{eq:eKS_anis}) and the one obtained in Ref.~\onlinecite{Munoz-Garcia2006} will be further discussed in the following sections, for the case of normal ion incidence.

\subsection{Effective Equation for Normal Incidence}

Having as a reference experimental behaviors those reported in Refs.~\onlinecite{Kim2011} and \onlinecite{Kim2013}, in which an initial Au-prepatterned surface was further irradiated at normal incidence, we will focus here in such condition $\theta=0$. This implies\cite{Munoz-Garcia2008,Munoz-Garcia2009} $\alpha_{1x} = 0$, $\alpha_{2x} = \alpha_{2y} = \alpha_2$, and $\alpha_{3x} = \alpha_{3y} = \alpha_3$, and will allow us to isolate the effects purely due to anisotropies in surface diffusion. For this reason, we will moreover assume isotropic surface tension, namely, $\gamma_{2x}=\gamma_{2y}=\gamma_2$. As in Refs.~ \onlinecite{Kim2011} and \onlinecite{Kim2013}, we will also take $x$ and $y$ to be aligned with the substrate directions along which surface diffusivities are optimized. Under these conditions, both the excavation and the addition rates become isotropic, Eq.\ (\ref{eq:eKS_anis}) taking the simpler form
\begin{eqnarray} \label{eq:ec_normal_anis2}
	\frac{\partial h}{\partial t} &=& -\nu \nabla^2 h + \lambda^{(1)} ( \nabla h )^2 \\ \nonumber & & - \nabla \cdot \left[ \boldsymbol{\mathcal{K}} \nabla \left( \nabla^2 h \right)\right] - \nabla \cdot \left\{\boldsymbol{\Lambda_2} \nabla  \left[ ( \nabla h )^2 \right]\right\},
\end{eqnarray}
where $\boldsymbol{\mathcal{K}}$ and $\boldsymbol{\Lambda_2}$ are matrices defined as
\begin{eqnarray}
	\boldsymbol{\mathcal{K}} =
	\left[
	\begin{array}{cc}
		\mathcal{K}_x & 0 \\
		0 & \mathcal{K}_y
	\end{array}
	\right] \mbox{\space and \space}
	\boldsymbol{\Lambda_2} =
	\left[
	\begin{array}{cc}
		\lambda^{(2)}_x & 0 \\
		0 & \lambda^{(2)}_y
	\end{array}
	\right]. \label{eq:K_and_Lambda}
\end{eqnarray}
The number of independent parameters in Eqs.\ \eqref{eq:ec_normal_anis2} and \eqref{eq:K_and_Lambda} has reduced dramatically, the remaining ones being $\nu = \epsilon  \phi  \gamma_0 \alpha_2$, $\lambda^{(1)} = -\epsilon \phi \gamma_0  \alpha_3$, $\mathcal{K}_i = D_i R_{eq} \gamma_2 + \epsilon ( \phi R_{eq} \gamma_0 \gamma_2 - \bar{\phi} D_i ) \alpha_2$, and $\lambda_i^{(2)} = \epsilon (  \phi R_{eq} \gamma_0 \gamma_2 - \bar{\phi} D_i ) \alpha_3$, where $i=x,y$. It is important to note that, in contrast to the equation obtained in Ref.~\onlinecite{Munoz-Garcia2006,Munoz-Garcia2009}, in which only terms of the form $\nabla^2 h$ and $(\nabla h )^2$ appear under normal ion incidence, in the case of Eq.\ \eqref{eq:ec_normal_anis2} the second-order derivatives $\partial^2/\partial x^2$ and $\partial^2/\partial y^2$ are weighted by parameters that depend on the different diffusion coefficients.

Under experimental conditions leading to pattern formation, $\nu>0$ in Eq.\ (\ref{eq:ec_normal_anis2}).
With respect to the coefficients of the linear fourth-order derivative term in this equation, note that they contain contributions that couple different physical mechanisms in a natural way. Thus, the contribution proportional to surface diffusivity and surface tension is completely analogous to the form of Mullins' surface diffusion, although note that the coefficients may also include ion-induced contributions which are temperature-independent.\cite{Munoz-Garcia2008} The remaining term in ${\cal K}_i$ couples erosion (being proportional to $\alpha_2$) with transport (surface diffusivity) and surface tension, further implementing ion-induced diffusivity. We will consider conditions in which this fourth-order derivative term has a net smoothing effect, so that $\mathcal{K}_i >0$. Finally, we consider the products $\lambda^{(1)} \lambda^{(2)}_{j}$ to also be positive. Mathematically, this condition is required for Eq.\ (\ref{eq:ec_normal_anis2}) to be free of so-called ``cancellation modes'', known to occur under appropriate conditions in related continuum models, such as the anisotropic KS \cite{Rost1995} and eKS \cite{Castro2005b,Munoz-Garcia2006} equations.

We next rescale Eq.\ (\ref{eq:ec_normal_anis2}) in order to work in dimensionless units. This allows to perform generic statements on the system behavior, while at the same time it also simplifies the discussion by minimizing the number of free parameters. Hence, we define
\begin{equation*}\label{eq:rescal2}
x = \Big( \frac{\mathcal{K}}{\nu} \Big)^{\frac{1}{2}} x',\; y = \Big( \frac{\mathcal{K}}{\nu} \Big)^{\frac{1}{2}} y',\;
t = \frac{\mathcal{K}}{\nu^2} t',\; h = \frac{\nu}{\lambda^{(1)}} h',
\end{equation*}
where $\mathcal{K} = \left({\mathcal{K}_x + \mathcal{K}_y}\right)/{2}$.
Dropping the primes, Eq.\ \eqref{eq:ec_normal_anis2} now reads
\begin{eqnarray}\label{eq:ec_normal_anis4}
	\frac{\partial h}{\partial t} &=& - \nabla^2 h + ( \nabla h )^2 - 2\nabla \cdot \left[ \boldsymbol{A} \nabla \left( \nabla^2 h \right)\right] \\ \nonumber & & -  2r_0 \nabla \cdot \left\{\boldsymbol{B}\nabla \left[( \nabla h )^2 \right] \right\} \mbox{\space ,}
\end{eqnarray}
where the matrices $\boldsymbol{A}$ and $\boldsymbol{B}$ are
\begin{align*}
	\boldsymbol{A} &= \frac{\boldsymbol{\mathcal{K}}}{\mathcal{K}_x + \mathcal{K}_y}=
		\left[
	\begin{array}{cc}
		\alpha_x & 0 \\
		0 & 1 - \alpha_x
	\end{array}
	\right], \\
	\boldsymbol{B} &= \frac{ \boldsymbol{\Lambda_2}}{\lambda_x^{(2)} + \lambda_y^{(2)}}=
	\left[
	\begin{array}{cc}
		\beta_x & 0 \\
		0 & 1 - \beta_x
	\end{array}
	\right],
\end{align*}
with
\begin{equation*}
	\alpha_x = \frac{ \mathcal{K}_x}{\mathcal{K}_x + \mathcal{K}_y}, \;
	\beta_x = \frac{\lambda_x^{(2)}}{\lambda_x^{(2)} + \lambda_y^{(2)}}, \;
	r_0 = \frac{\nu}{\lambda^{(1)}} \Big( \frac{\lambda_x^{(2)} + \lambda_y^{(2)}}{\mathcal{K}_x + \mathcal{K}_y} \Big).
\end{equation*}
Given that three independent rescalings have been performed on Eq.\ \eqref{eq:ec_normal_anis2}, which depends on six independent parameters, the final Eq.\ \eqref{eq:ec_normal_anis4} depends on three independent constants only, $\alpha_x$, $\beta_x$, and $r_0$. Note that, in all the physically relevant cases, $r_0>0$.

It is interesting to stress some of the features of Eq.\ \eqref{eq:ec_normal_anis4}: {\em (i)} As expected, the anisotropies are only caused by the different diffusivities $D_x$ and $D_y$. This is reflected in the fact that the parameters $\alpha_x$ and $\beta_x$ will generally take values different from 1/2. Therefore, the weights for both directions in the last two terms of Eq.\ \eqref{eq:ec_normal_anis4} will be different. When $\alpha_x = \beta_x = 1/2$ the isotropic diffusion equation for normal incidence proposed in Ref.\ \onlinecite{Castro2005} is recovered.  {\em (ii)} The dimensionless parameter $r_0$ is the squared ratio of two length scales. One of these length scales is computed as the ratio between the parameters of the conserved Kardar-Parisi-Zhang (KPZ) \cite{Munoz-Garcia2006} nonlinear terms appearing in the equation, $\lambda_i^{(2)}$, and that of the non-conserved KPZ nonlinearity that ensues, $\lambda^{(1)}$, namely, $[(\lambda_x^{(2)} + \lambda_y^{(2)}]/\lambda^{(1)})^{1/2}$. The second length scale is set by the parameters of the linear terms, as $[(\mathcal{K}_x + \mathcal{K}_y)/\nu]^{1/2}$. The parameter combination $r_0$ thus provides an estimate of the relative relevance of the various contributions that compete in the dynamics of the system. 

As just stressed, in general Eq.\ \eqref{eq:ec_normal_anis4} is anisotropic, thus the patterns will present different wavelengths along each principal direction. These can be estimated as functions of the parameters of the equation, by performing a standard linear stability analysis.\cite{Munoz-Garcia2008} This leads to
\begin{equation}\label{eq:Lxy_anis}
	\ell_i = \frac{2 \pi} {k_i^{\ell}} = 2 \pi \sqrt{\frac{2 \mathcal{K}_i}{\nu}}\approx 2\pi \sqrt{\frac{2 R_{eq} D_{i} \gamma_{2i}}{\epsilon \phi \gamma_0 \alpha_{2 i}}},
\end{equation}
where we have substituted the values of $\mathcal{K}_i$ and ${\nu}$ provided after Eq.\ \eqref{eq:K_and_Lambda}. As expected, these wavelengths coincide with the values obtained in Section \ref{Model} for the linear stability analysis of the full two-field model. Importantly, the expressions obtained for $\ell_i$ can be often used to perform (semi)quantitative comparisons between the present type of continuum models and experiments at short times, prior to the onset of non-linear effects.\cite{Munoz-Garcia2010,Munoz-Garcia2012,Kim2013}

\section{Results}\label{sec:simulation}

Thus far, we have been able to derive an effective dimensionless equation for normal ion incidence that contains all the physical mechanisms of the problem and depends on three free parameters only, Eq.\ \eqref{eq:ec_normal_anis4}. In this section we study systematically this equation by independently changing the values of each of these three parameters. Given the strong nonlinearities in the equation, we resort to a numerical integration. Specifically, our code was implemented in MATLAB, being based on a standard finite-difference scheme in space (for the linear terms) and a fourth-order Runge-Kutta method for the time evolution, using a spatial grid with 256$\times$256 nodes, a time step $\Delta t = 0.01$, and a space step $\Delta x =1$. The discretization of the nonlinear terms was based on the one proposed by Lam and Shin in Ref.~\onlinecite{Lam1998}. We have employed periodic boundary conditions and initial height values which are uniformly distributed between $0$ and $0.1$. Besides inspection of the resulting surface morphologies, in all cases we have calculated the global surface roughness, $W$, as well as the wavelengths, $\ell_x$ and $\ell_y$, after averaging over 10 realizations of the initial condition for each parameter set. Additionally, we have computed the normalized autocorrelation function, $R_N$, that allows one to determine the local arrangement of the patterns and is defined as\cite{Zhao}
\begin{equation}\label{eq:Ra_auto}
	 R_N (\boldsymbol{x}, t) =  \frac{1}{W} \frac{1}{L^2} \int [h (\boldsymbol{x} + \boldsymbol{r}, t) h (\boldsymbol{r}, t) - \bar{h}^2 (t) ] d \boldsymbol{r} ,
\end{equation}
where $W$ is the surface roughness and $\bar{h}$ is the mean height over the whole spatial grid of size $L \times L$.

\subsection{Isotropic case: $\alpha_x = \beta_x = 1/2$}\label{sec:simulation_isotropic}
To begin with our analysis, and for the sake of later comparison with anisotropic parameter conditions, we first recall the results obtained in Ref.\ \onlinecite{Munoz-Garcia2009} for isotropic systems under normal ion incidence, as a special case of our model in which both  surface diffusivities are equal, $D=D_x = D_y$. In this case Eq.\ \eqref{eq:ec_normal_anis2} simply reduces to
\begin{equation}\label{eq:eKS_normal}
	\frac{\partial h}{\partial t} = -\nu \nabla^2 h - \mathcal{K} \nabla^4 h  + \lambda^{(1)} ( \nabla h )^2 - \lambda^{(2)} \nabla^2 ( \nabla h )^2,
\end{equation}
where $\mathcal{K} = \mathcal{K}_x = \mathcal{K}_y$ and $\lambda^{(2)} = \lambda_x^{(2)} = \lambda_y^{(2)}$. After a rescaling which is similar to the one employed in the previous section, this equation reduces to the particular case of Eq.\ \eqref{eq:ec_normal_anis4} in which $\alpha_x = \beta_x = 1/2$. Note that, in principle, the simulations reported in Ref.\ \onlinecite{Munoz-Garcia2009} correspond to the unrescaled Eq.\ (\ref{eq:eKS_normal}).

The following features for the surface roughness and pattern wavelength were obtained in this case:\cite{Munoz-Garcia2009} {\em (i)} The surface morphology shows a short-time transient behavior. During this interval $W$ grows exponentially with time and a dot pattern appears whose characteristic wavelength is accurately described by the linear analysis. Indeed, this stage is controlled by the linear terms $\nu \nabla^2 h$ and $\mathcal{K} \nabla^4 h$. {\em (ii)} After this linear regime, a crossover takes place towards a behavior which is controlled by the conserved nonlinear term $\lambda^{(2)} \nabla^2 ( \nabla h )^2$, in which the growth of $W$ and $\ell$ in time can be approximated by power-laws, with effective exponents whose values depend on equation parameters. {\em (iii)} For long times, the non-conserved nonlinear term $\lambda^{(1)} ( \nabla h )^2$ induces eventual saturation of $W$ and $\ell$, and height disorder at large scales.

Actually, the relative duration of the various dynamical regimes turns out to be controlled by the parameter $r_0$.\cite{Munoz-Garcia2006,Munoz-Garcia2009} Thus, large $r_0$ values correspond to the predominance of the conserved KPZ nonlinearity at intermediate times, allowing for a stronger coarsening process and an improved order of the height values throughout the surface, namely, a smaller roughness. On the contrary, small $r_0$ values correspond to a non-linear regime dominated by the KPZ nonlinearity, with a relatively short intermediate coarsening regime and a rougher surface at long times.

\begin{figure*}[!htbp]
\centering
\subfloat[]
{ \includegraphics[width=9cm]{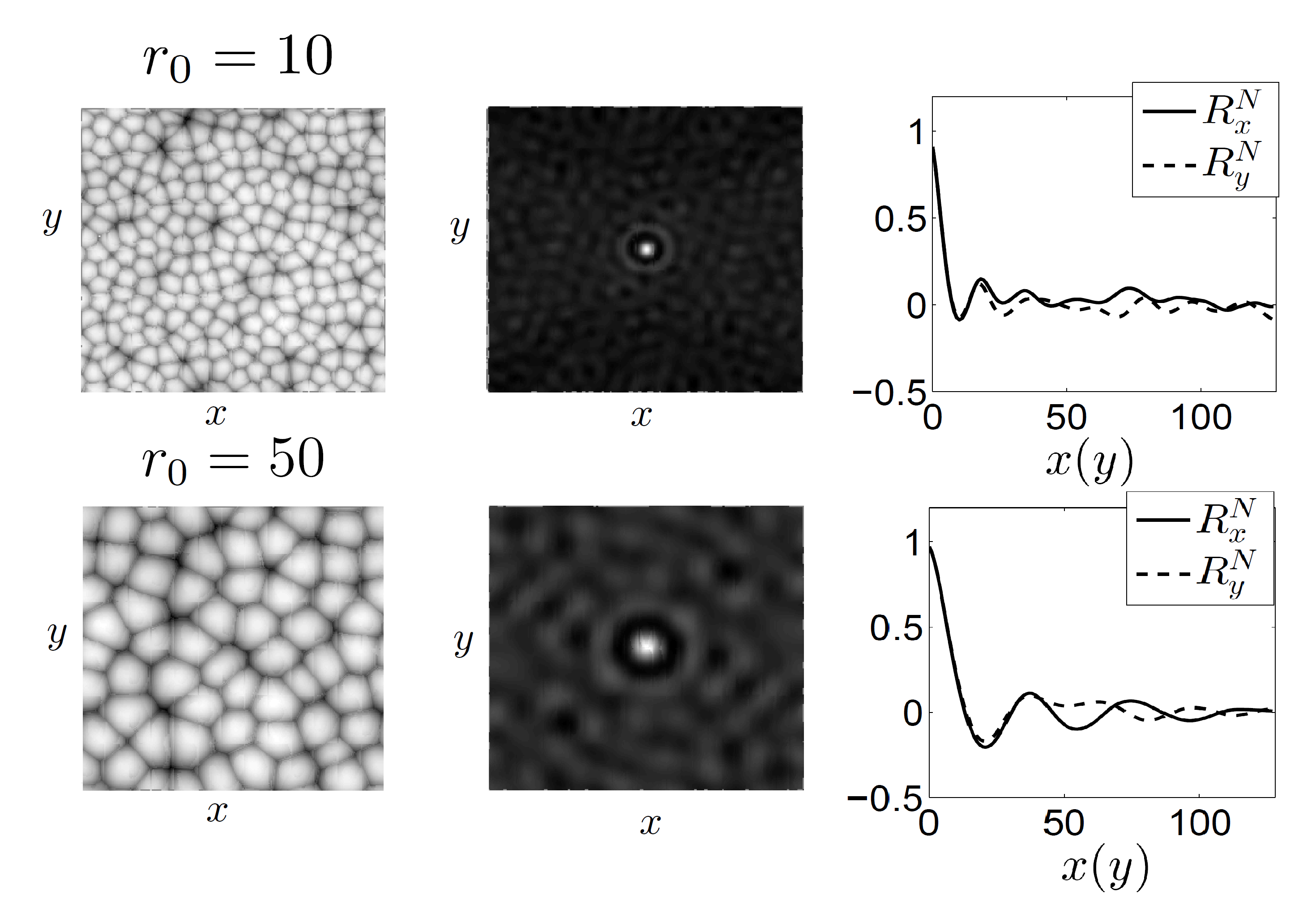}
\label{fig:Fig0a}
}
\subfloat[]
{ \includegraphics[width=5.75cm]{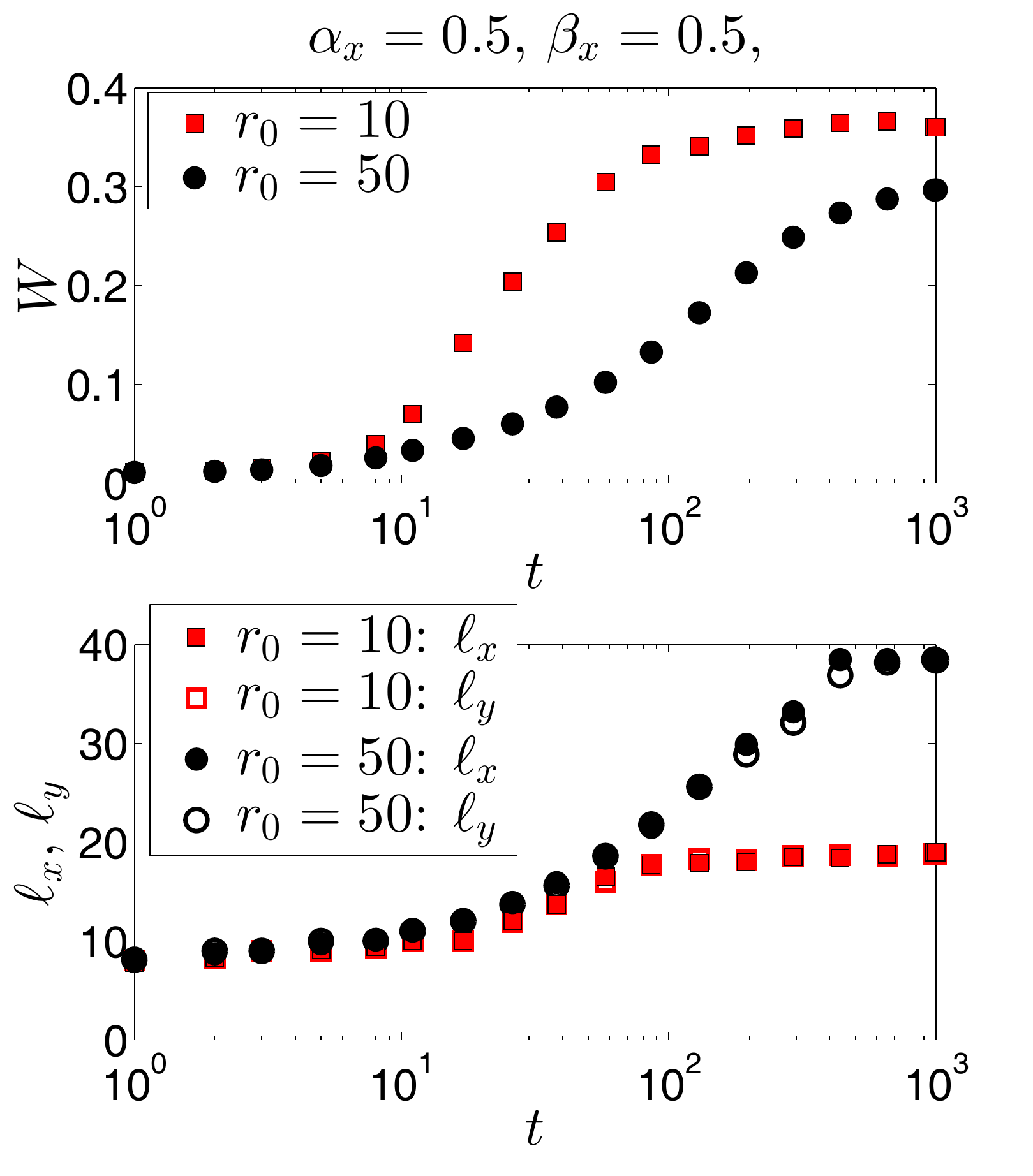}	
\label{fig:Fig0b}
}
\caption{(Color online) (a) Top-view surface morphologies (left column), normalized autocorrelation functions, $R_N$ (center column), and normalized autocorrelation functions along the $x$ and $y$ directions, $R_x^N$ and $R_y^N$ respectively (right column), predicted by Eq.\ (\ref{eq:ec_normal_anis4}) at $t=1000$ for $\alpha_x=0.5$, $\beta_x=0.5$, and different values of $r_0$ (see legends). (b) Temporal evolution of the roughness, $W$, and the wavelengths along the $x$ and $y$ directions, $\ell_x$ (solid symbols) and $\ell_y$ (open symbols) respectively, for the same parameter values as in (a).}
\label{fig:iso}
\end{figure*}

At this point, it is important to remark that the 1D and 2D behaviors of Eq.\ \eqref{eq:eKS_normal} differ quite strongly with respect to the ordering properties. To begin with, note that, given the fact that the band of unstable modes extends down to $\mathbf{k}=\mathbf{0}$, combined with the occurrence of the non-conserved KPZ nonlinearity, order in the pattern can be short-range at most.\cite{Cross2009} This does not prevent the equation from providing a quantitatively accurate description of experimental patterns.\cite{Munoz-Garcia2010,Munoz-Garcia2012} Then, while relative homogeneity in height values correlates positively with an enhanced ``in-plane'' order for Eq.\ \eqref{eq:eKS_normal} in 1D,\cite{Munoz-Garcia2006} this is not the case in 2D. Namely, for the 2D case, smaller $r_0$ values
seem to feature improved in-plane short-range hexagonal ordering of the dot structure. Conversely, larger values of $r_0$, that allow for stronger coarsening (wider cells) and smaller overall roughness, correspond to surfaces with poorer in-plane ordering. For a thus-far unreported explicit comparison, see Fig.\ \ref{fig:iso}, in which the $r_0=10$ and 50 cases are explicitly illustrated. The results shown in Figure \ref{fig:iso} also show that in both cases the wavelengths along each direction, $\ell_x$ and $\ell_y$, are equal, since the surface diffusion is isotropic.

The rich crossover behavior of Eq.~\eqref{eq:eKS_normal} will be useful to better understand the morphologies described by Eq.~\eqref{eq:ec_normal_anis4} as a function of parameters values.

\subsection{General values of $\alpha_x$}\label{sec:simulation_alpha}

To study the effect of non-isotropic values of the parameter $\alpha_x$ mediating the linear surface-diffusion terms in Eq.\ (\ref{eq:ec_normal_anis4}), we have integrated numerically Eq.\ \eqref{eq:ec_normal_anis4} for $r_0=10$, $\beta_x=0.5$, and values of $\alpha_x \in [0,\frac{1}{2}]$. Due to the symmetry of Eq.\ \eqref{eq:ec_normal_anis4} with respect to reflections of $\alpha_x$ around the isotropic 1/2 value, the behavior for $\alpha_x \in [\frac{1}{2}, 1]$ can be easily obtained from our simulations by simply swapping the $x$ and $y$ axes.

\begin{figure*}[!htbp]
\centering
\subfloat[]
{ \includegraphics[width=9cm]{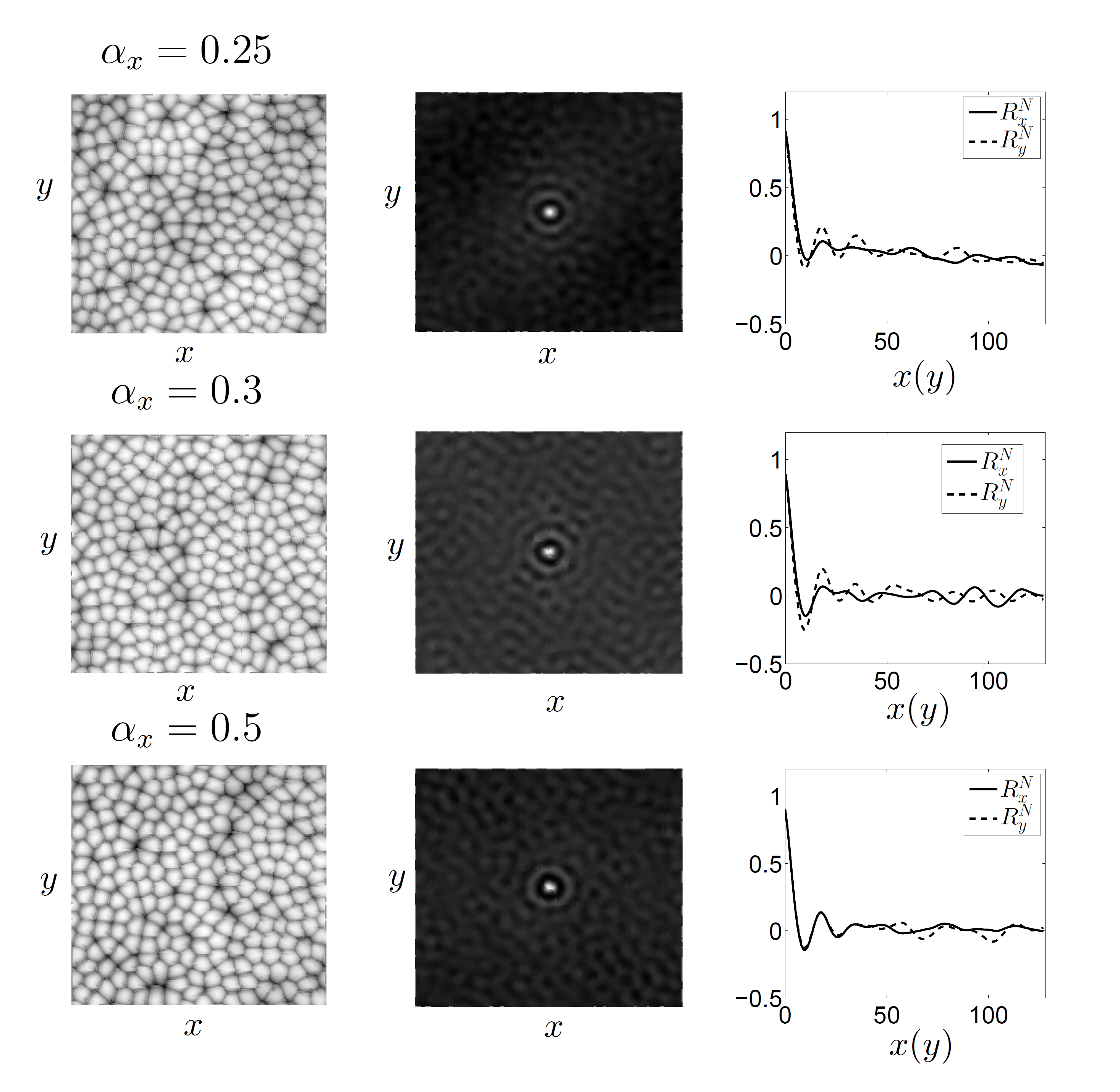}
\label{fig:Fig1a}
}
\subfloat[]
{ \includegraphics[width=5.75cm]{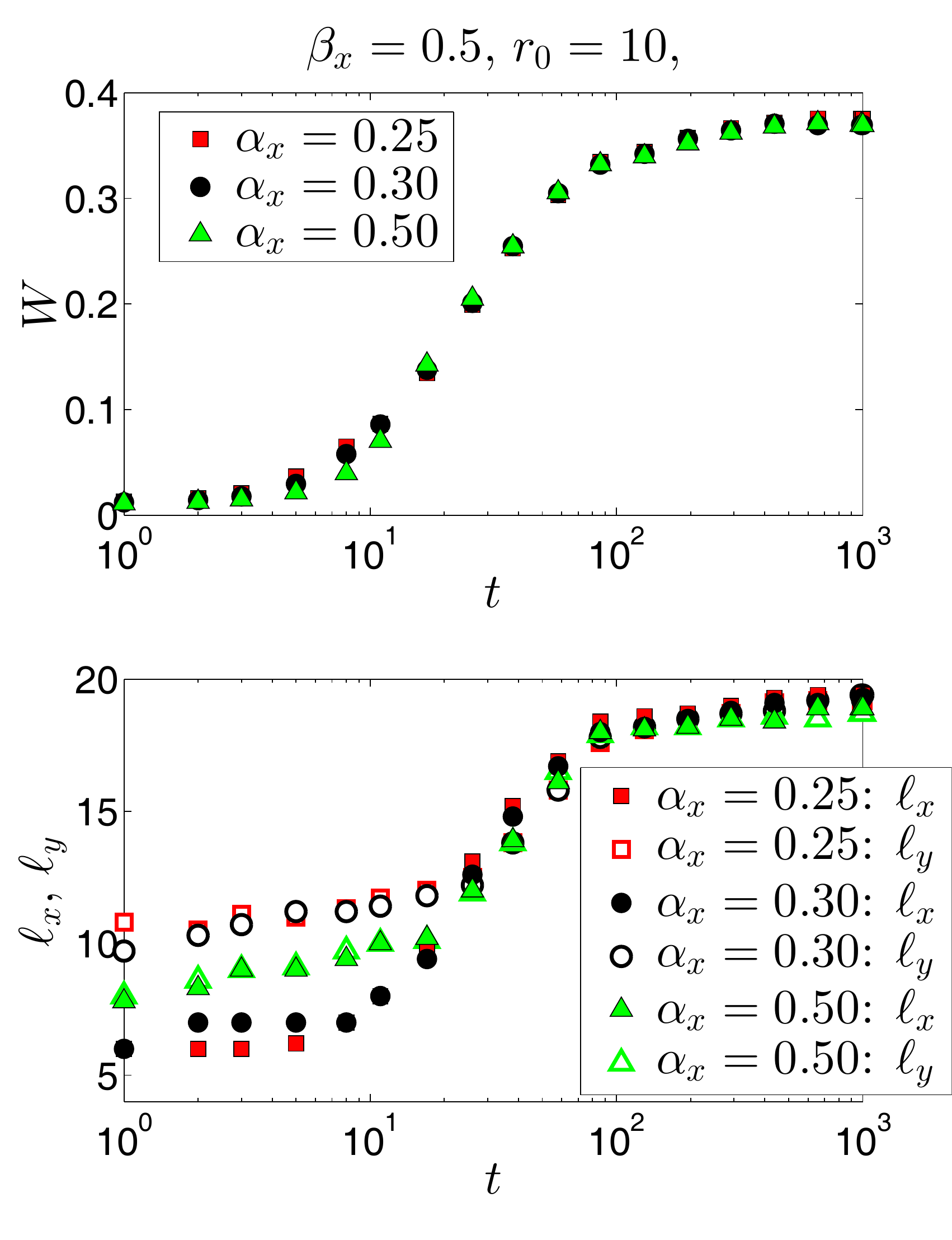}
\label{fig:Fig1b}
}
\caption{(Color online) (a) Top-view surface morphologies (left column), normalized autocorrelation functions, $R_N$ (center column), and normalized autocorrelation functions along the $x$ and $y$ directions, $R_x^N$ and $R_y^N$ respectively (right column), predicted by Eq.\ (\ref{eq:ec_normal_anis4}) at $t=1000$ for $\beta_x=0.5$, $r_0=10$, and different values of $\alpha_x$ (see legends). (b) Temporal evolution of the roughness, $W$, and the wavelengths along the $x$ and $y$ directions, $\ell_x$ (solid symbols) and $\ell_y$ (open symbols) respectively, for the same parameter values as in (a).}%
\label{fig:H_alpha}
\end{figure*}

Figure \ref{fig:Fig1a} shows the surface morphologies, the normalized autocorrelation function, and cross-cuts of the normalized autocorrelation function along the $x$ and $y$ axes, $R_x^N$ and $R_y^N$, respectively, at $t=1000$ and for different values of $\alpha_x$. Note that, as discussed above, $\alpha_x=0.5$ corresponds to the isotropic case already studied in Ref.\ \onlinecite{Munoz-Garcia2009}. As noticed by inspecting the left column of the figure, the surface morphology does not change qualitatively for different values of $\alpha_x$. This {\em robustness} with respect to $\alpha _x$ is further evidenced in the middle and right panels of Fig.~\ref{fig:Fig1a}, in which $R^N$ is shown side-by-side with $R_x^N$ and $R_y^N$. The $x$ and $y$ wavelengths, $\ell_x$ and $\ell_y$, can be obtained by measuring the distance from the origin to the first maximum of the autocorrelation function along the corresponding axis. For all the values of $\alpha_x$ considered, we obtain $\ell_x \simeq \ell_y$, the only significant difference being that, for $\alpha_x=0.25$ and $\alpha_x=0.3$, the first peak of the autocorrelation function along the $y$-direction is higher than the first peak along the $x$-direction, both peaks having the same heights for $\alpha_x=0.5$. This implies that, for $\alpha_x \in (0, 0.5)$, dots are more correlated along the $y$-direction, although the differences are not substantial.

Figure \ref{fig:Fig1b} shows the time evolution of the global roughness $W$ and wavelengths, $\ell_x$ (solid symbols) and $\ell_y$ (open symbols), for the same values of $\alpha_x$ as in Fig. \ref{fig:Fig1a}. As for the isotropic eKS model, three time regimes can be distinguished: Initially the roughness grows exponentially, up to intermediate times when its growth rate slows down; eventually it reaches a similar time-independent value for all $\alpha_x$. On the other hand, the behavior of the pattern wavelengths with $\alpha_x$ is different. Initially both $\ell_x$ and $\ell_y$ start growing slowly, with $\ell_y$ being larger than $\ell_x$ for $\alpha_x<0.5$. This is due to the fact that the fourth-order linear terms controlled by the parameter $\alpha_x$ are expected to play an important role precisely at the small spatial and time scales at which the linear instability develops. As a matter of fact, looking at the expressions of the wavelengths predicted by the linear instability analysis, Eq.\ \eqref{eq:Lxy_anis}, we can easily note that $\ell_x$ should be smaller than $\ell_y$ if $\mathcal{K}_x<\mathcal{K}_y$, which is indeed the case for $\alpha_x<0.5$. This linear transient behavior is followed by a coarsening process controlled by the nonlinear terms, which finally drive both wavelengths to similar saturation values. Thus, since the parameters of the nonlinear terms $r_0$ and $\beta_x$ are fixed, the final surface topographies are very similar at long times for the different values of $\alpha_x$. Additionally, because the nonlinear terms are isotropic ($\beta_x = 0.5$) the wavelengths reach similar values in both directions at long times. In the next sections we study the impact of the coefficients of the nonlinear terms on the system dynamics and pattern formation and evolution.


\subsection{General values of $\beta_x$}\label{sec:simulation_beta}

\begin{figure*}[!htbp]
\centering
\subfloat[]
{ \includegraphics[width=9cm]{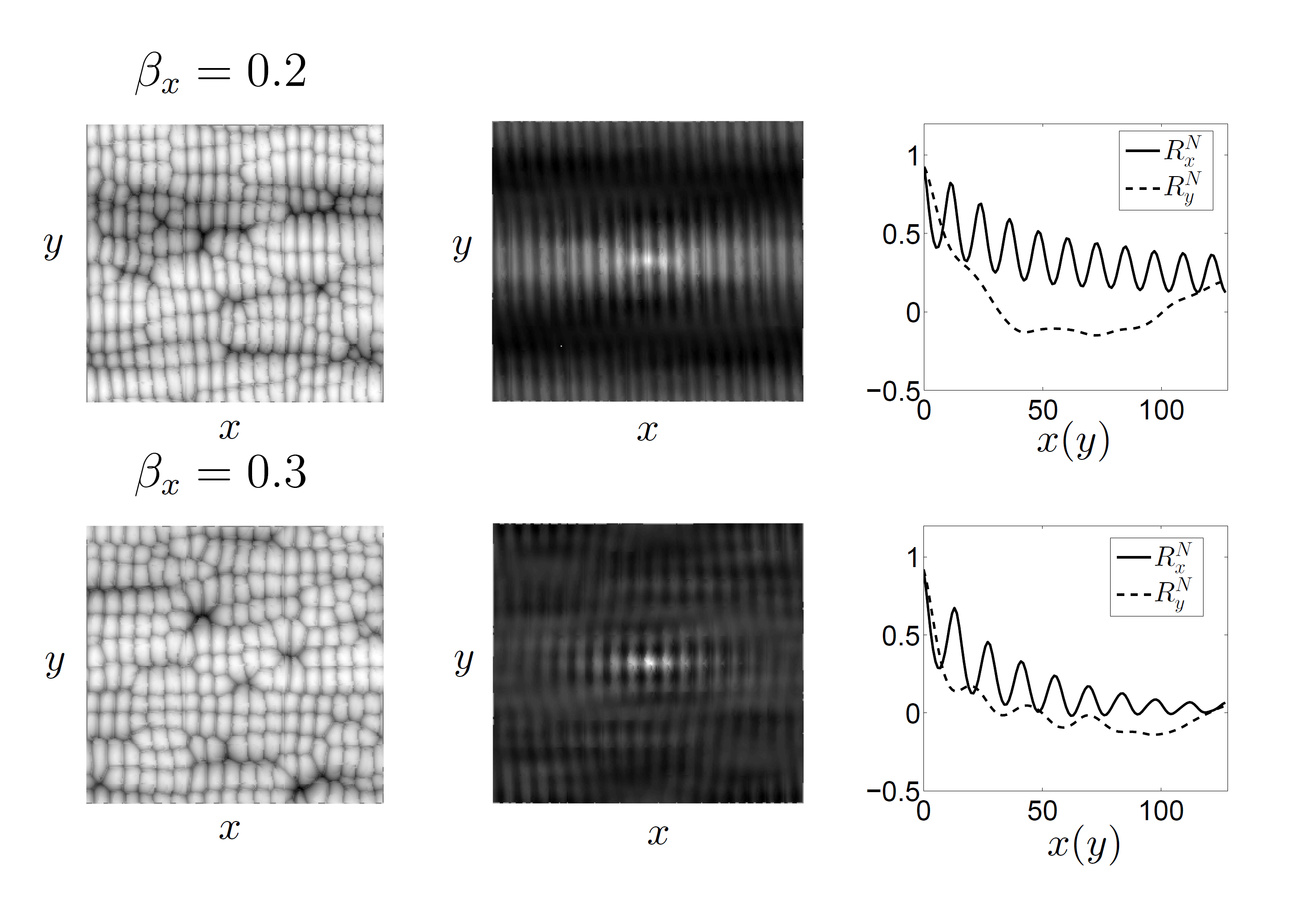}
\label{fig:Fig2a}
}%
\subfloat[]
{ \includegraphics[width=5.75cm]{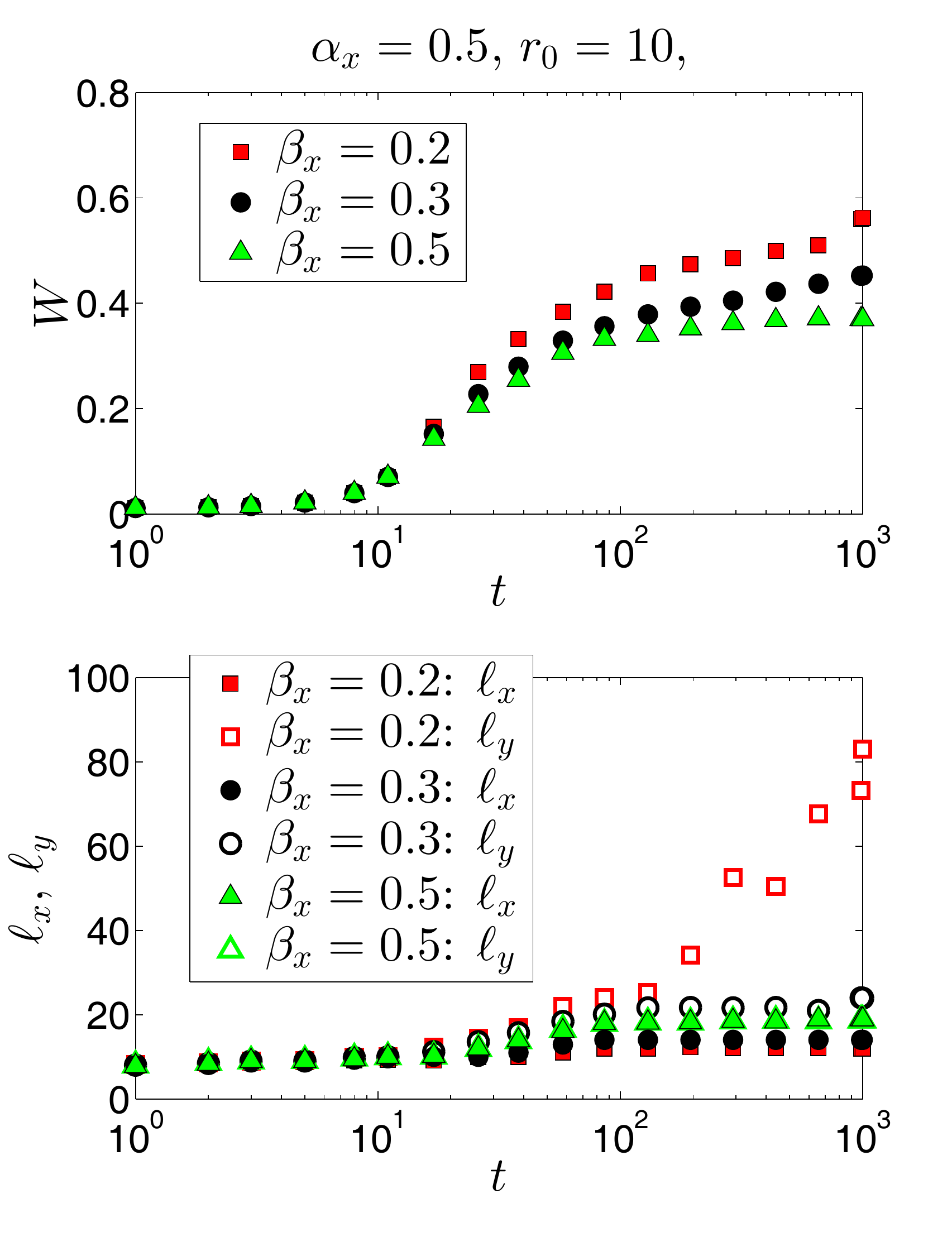}
\label{fig:Fig2b}
}
\caption{(Color online) (a) Top-view surface morphologies (left column), normalized autocorrelation functions, $R_N$ (center column), and normalized autocorrelation functions along the $x$ and $y$ directions, $R_x^N$ and $R_y^N$ respectively (right column), predicted by Eq.\ (\ref{eq:ec_normal_anis4}) at $t=1000$ for $\alpha_x=0.5$, $r_0=10$, and different values of $\beta_x$ (see legends). (b) Temporal evolution of the roughness, $W$, and the wavelengths along the $x$ and $y$ directions, $\ell_x$ (solid symbols) and $\ell_y$ (open symbols) respectively, for the same parameter values as in (a).}%
\label{fig:H_beta}
\end{figure*}

We next consider the influence on the topography of the anisotropic, conserved nonlinearity which is controlled in Eq.\ (\ref{eq:ec_normal_anis4}) by the parameter $\beta_x$. To this end,
numerical integrations of Eq.\ \eqref{eq:ec_normal_anis4} have been performed for fixed values of $\alpha_x$ and $r_0$, and $\beta_x \in [0,\frac{1}{2}]$. Similarly to the case of $\alpha_x$, results for $\beta_x \in [\frac{1}{2}, 1]$ can be deduced from the simulations shown next by swapping the $x$ and $y$ axes. Figure \ref{fig:Fig2a} displays the morphology, the normalized autocorrelation function, and the autocorrelation function along the $x$ and $y$ axes at $t=1000$ for different values of $\beta_x$. The additional case $\beta_x=0.5$ for the chosen $\alpha_x$ corresponds to the isotropic system already shown on the third row of Fig.\ \ref{fig:Fig1a}.
The temporal evolution of the surface roughness and wavelengths are shown in figure \ref{fig:Fig2b}.

If $\beta_x$ is small,
the conserved nonlinearity acts predominantly along the $y$-axis, inducing stronger coarsening behavior, hence $\ell_y$ becomes larger than $\ell_x$ for times beyond the linear regime, see Fig.\ \ref{fig:Fig2b}. Actually, this behavior is associated with a change in the pattern symmetry, see e.g.\ the $\beta_x=0.2$ case in Fig.\ \ref{fig:Fig2a}. Indeed, the larger value of $\ell_y$ implies that the dots or cells become more elongated in the $y$-direction, leading to the emergence of a ripple pattern with ridges parallel to it. This can be noted both in the morphology and in the autocorrelation functions, and is in spite of the fact that we are considering normal incidence conditions for the ions. Note, this is a purely non-linear effect, as Eq.\ (\ref{eq:ec_normal_anis4}) is completely isotropic {\em at linear order} for this parameter condition. On the other hand, if $\beta_x$ increases, the elongation of the dots along the $y$-direction is attenuated, they form arrangements with a more square (rather than rectangular) symmetry, and the effect is mitigated, see Fig.\ \ref{fig:Fig2a} for
$\beta_x = 0.3$. At any rate, for $\beta_x\neq 1/2$ the isotropy of the pattern is clearly broken.

With respect to the time evolution of the surface roughness, Fig.\ \ref{fig:Fig2b} indicates an unambiguous dependence with the value of $\beta_x$, which contrasts with the results obtained for $\alpha_x$ in the preceding section. For small values of $\beta_x$, saturation occurs later and the saturation value is larger. This seems reminiscent of results for the 1D eKS equation when increasing the strength of the conserved KPZ nonlinearity with respect to the remaining terms in the equation.\cite{Munoz-Garcia2006} On the other hand, for short times the roughness values are practically the same for all $\beta_x$, suggesting that such an increase of the roughness is  indeed a nonlinear effect. Regarding the pattern wavelengths in the two directions, both grow very slowly and take similar values during the short times associated with the linear instability. At intermediate times, both grow at increased rates; ultimately, they reach very different saturation values depending on the specific value of $\beta_x$.
Indeed, as already noted above, for relatively small values of this parameter the pattern wavelength in the $y$-direction, $\ell_y$, becomes larger than $\ell_x$, as can be clearly appreciated in Fig.\ \ref{fig:Fig2b} already for $\beta_x=0.3$. For even smaller values of $\beta_x$, such as $\beta_x=0.2$, $\ell_x$ interrupts its growth process early while $\ell_y$ keeps growing for a long time (note that its coarsening process has not yet stopped at $t=1000$ for $\beta_x=0.2$), resulting into very different $\ell_y>\ell_x$. This is due to the fact that the conserved nonlinear term, which induces the coarsening process, is stronger along the $y$ direction.
In summary, the role of $\beta _x$ is twofold: it modifies the local arrangement (symmetry and order) of the patterns and it amplifies/reduces the coarsening dynamics selectively along one of the system directions.

\subsection{General values of $r_0$}\label{sec:simulation_r}

\begin{figure*}[!htbp]
\centering
\subfloat[]
{ \includegraphics[width=9cm]{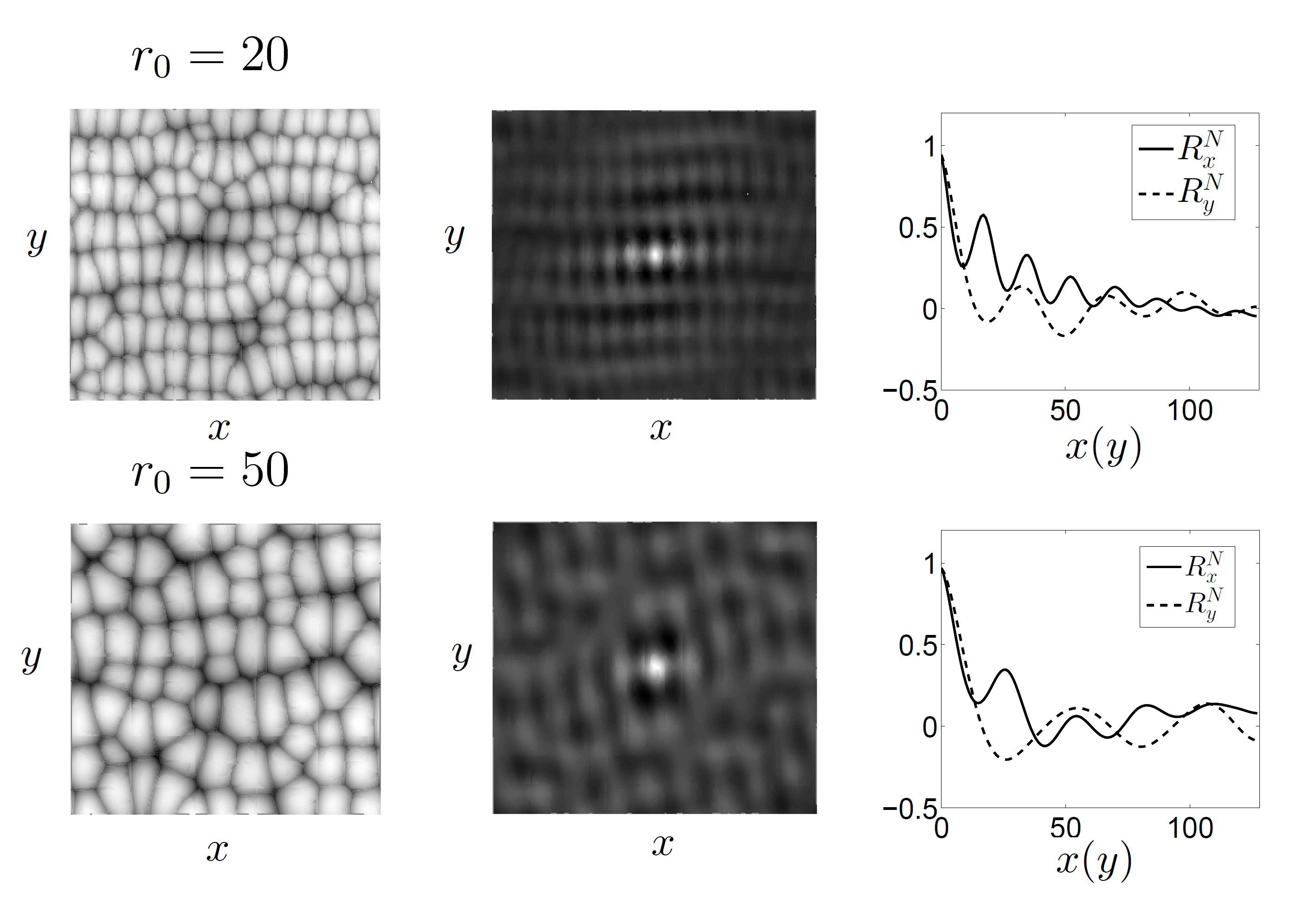}
\label{fig:Fig3a}
}%
\subfloat[]
{ \includegraphics[width=5.75cm]{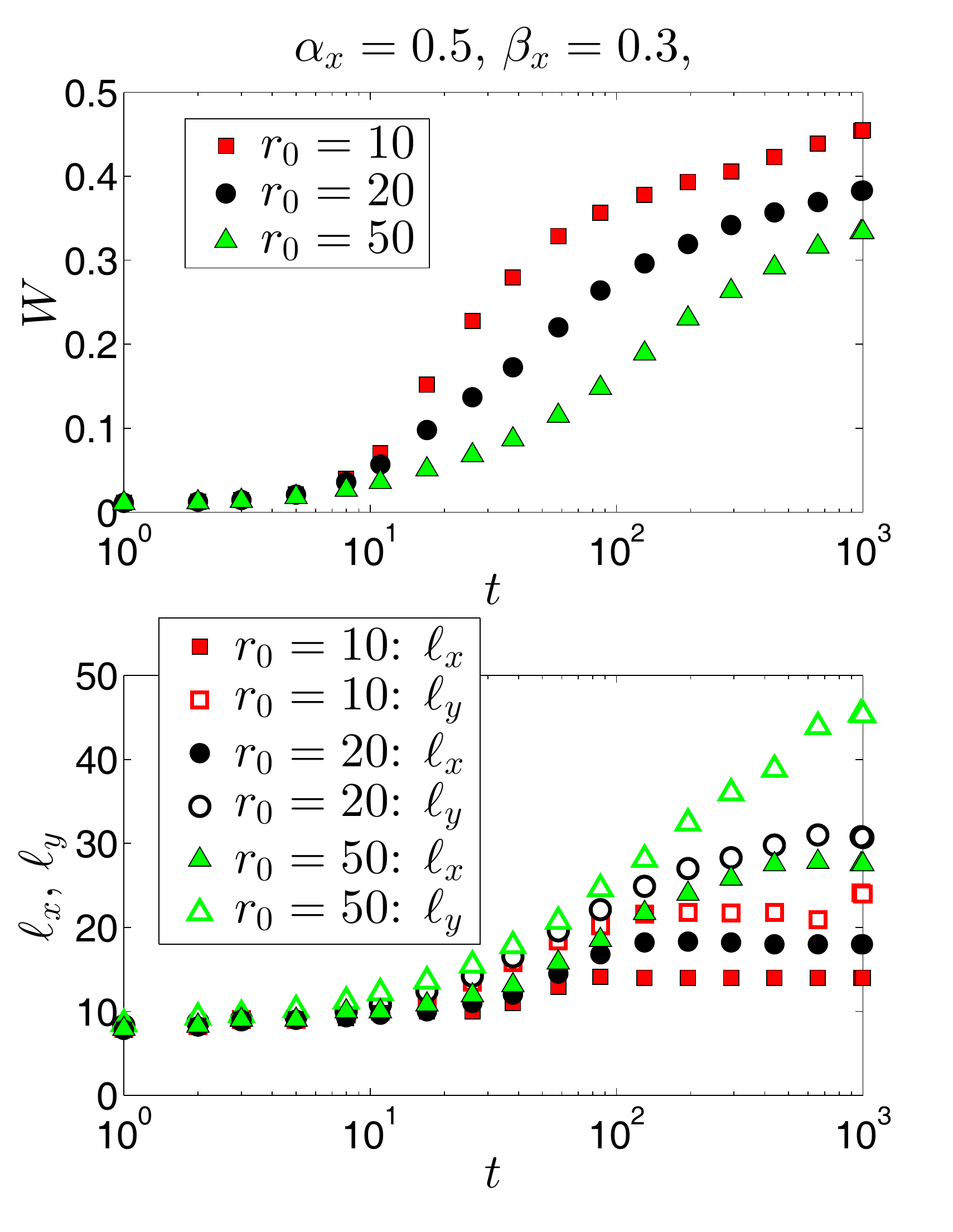}
\label{fig:Fig3b}
}
\caption{(Color online) (a) Top-view surface morphologies (left column), normalized autocorrelation functions, $R_N$ (center column), and normalized autocorrelation functions along the $x$ and $y$ directions, $R_x^N$ and $R_y^N$ respectively (right column), predicted by Eq.\ (\ref{eq:ec_normal_anis4}) at $t=1000$ for $\alpha_x=0.5$, $\beta_x=0.3$, and different values of $r_0$ (see legends). (b) Temporal evolution of the roughness, $W$, and the wavelengths along the $x$ and $y$ directions, $\ell_x$ (solid symbols) and $\ell_y$ (open symbols) respectively, for the same parameter values as in (a).}
\end{figure*}

We continue in this section with the morphological effects of the third independent parameter in Eq.\ \eqref{eq:ec_normal_anis4}, namely, the ratio of nonlinear to linear length scales, $r_0$. The simulation results for different values of $r_0$ are shown in Fig.\ \ref{fig:Fig3a}. Analogously to the isotropic case for normal incidence studied in Section \ref{sec:simulation_isotropic} and illustrated in Figs.\ \ref{fig:Fig0a} and \ref{fig:Fig0b}, in the presence of anisotropic surface diffusion the patterns present more coarsening and a smaller roughness when $r_0$ is larger. However, the quality of in-plane ordering of the dots is poorer.
For the parameter values considered in Fig.\ \ref{fig:Fig3a}, slightly elongated dots group together following square arrangements, as can be noted looking at the surface morphologies. However, the short-range square order is hindered for larger $r_0$ values. This is also reflected in the autocorrelation function, where a more perfect square pattern is revealed for smaller values of $r_0$.

The temporal evolution of the roughness and wavelengths for different values of $r_0$ are represented in Fig.\ \ref{fig:Fig3b}. Again three main regimes can be distinguished. The roughness grows exponentially in the first, linear regime, followed by power-law growth, and by saturation at very long times. As in Refs.~\onlinecite{Munoz-Garcia2006} and \onlinecite{Munoz-Garcia2009}, the final roughness is indeed smaller for large $r_0$ values, the long-time configurations showing more uniform height values. For such large $r_0$, the two wavelengths $\ell_x$ and $\ell_y$ are also larger, due to the longer coarsening process undergone. Note, because $\alpha_x=0.5$, the linear terms have the same effect in both directions.
Since $\beta_x=0.3$ in the simulations shown, and as we saw in the previous section, the wavelength grows more in the $y$ direction and patterns with $\ell_y>\ell_x$ are always obtained.


\subsection{Unusual patterns and order under normal incidence: Ripples and square or hexagonal dot arrangements}



As suggested by Fig.\ \ref{fig:Fig3a}, for intermediate values of $\beta_x$ it is possible to generate surfaces for which the dot patterns display short-range order with square symmetry. Such a morphology is locally characterized by each single dot having on average four nearest neighbors located along the two Cartesian directions.
Here we employ the normalized height autocorrelation function, $R_N$, to quantify the spatial order on the surface. If the morphology does correspond to a pattern with such a square-symmetric order, the central maximum of $R_N$ lies within a perfect square formed by eight nearest satellite peaks on a square arrangement. Although the surfaces described by Eq.\ \eqref{eq:ec_normal_anis4} present this type of structure to a certain degree, it is not possible to obtain a strictly square symmetry due to the anisotropy introduced by the conserved nonlinearities controlled by the parameter $\beta_x$. Indeed, the heterogeneous surface diffusivities in the two space directions lead to different wavelengths, even under isotropic (normal incidence) irradiation, producing a relatively ordered array of dots, but with different typical sizes in each direction. For $\beta \in (0, 0.5)$, dots are more correlated with their neighbors along the $x$ direction than along the $y$ direction, as can be noticed in the height autocorrelation functions obtained in the previous sections. See for example Fig.\ \ref{fig:Fig3a}, where the correlation values are clearly larger along the $x$-axis. Recall that decreasing $\beta_x$ in this range of values actually increases the elongation of dots along the $y$ axis.

\begin{figure}[!htmb]
\begin{center}
\begin{minipage}[c]{0.333\linewidth}
\begin{center}
{$\beta_x=0.1$}
\includegraphics[width=\linewidth]{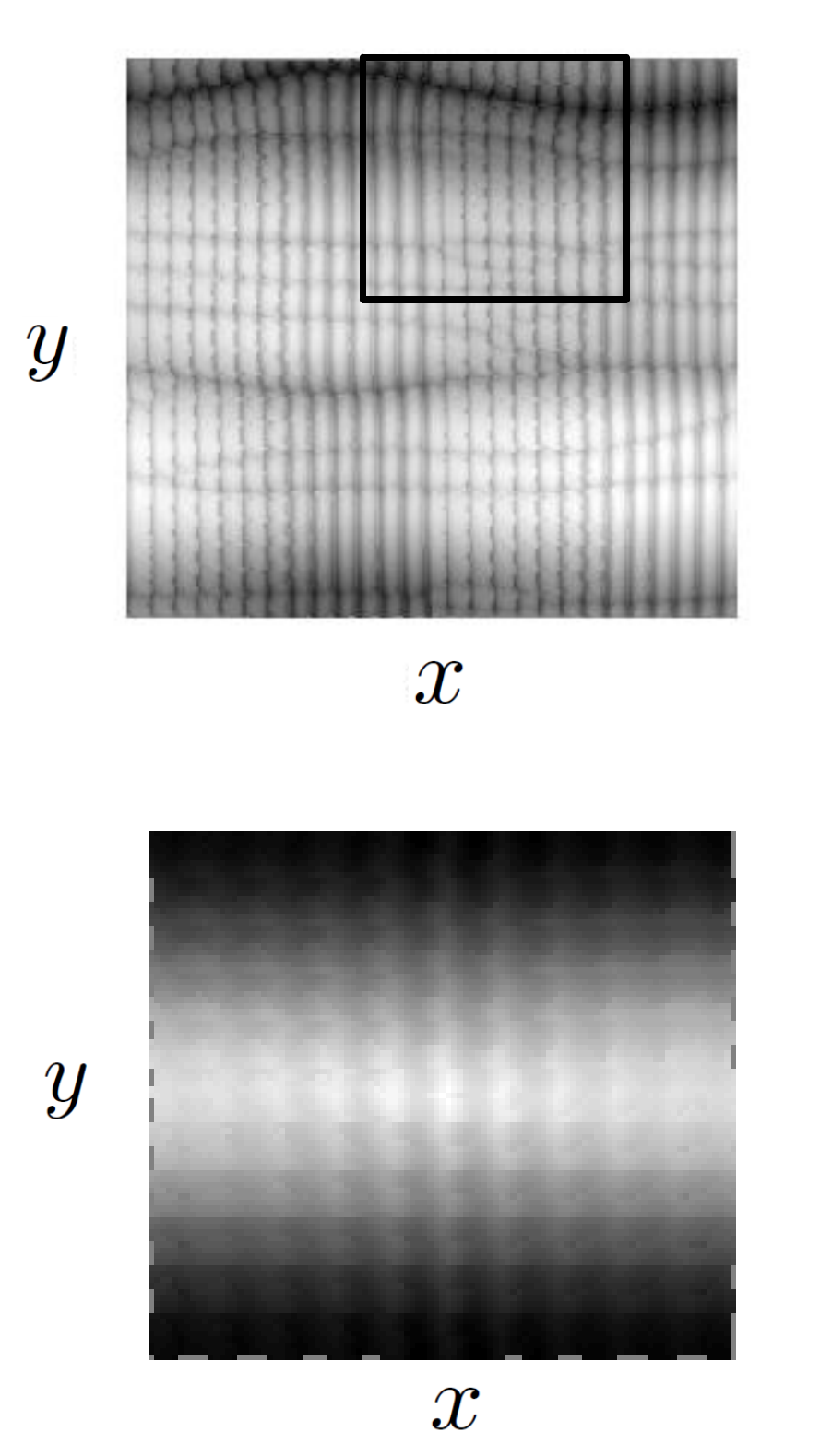}
\end{center}
\end{minipage}\hspace*{ 0.01\linewidth}
\begin{minipage}[c]{0.333\linewidth}
 \begin{center}
{$\beta_x=0.25$}
\includegraphics[width=\linewidth]{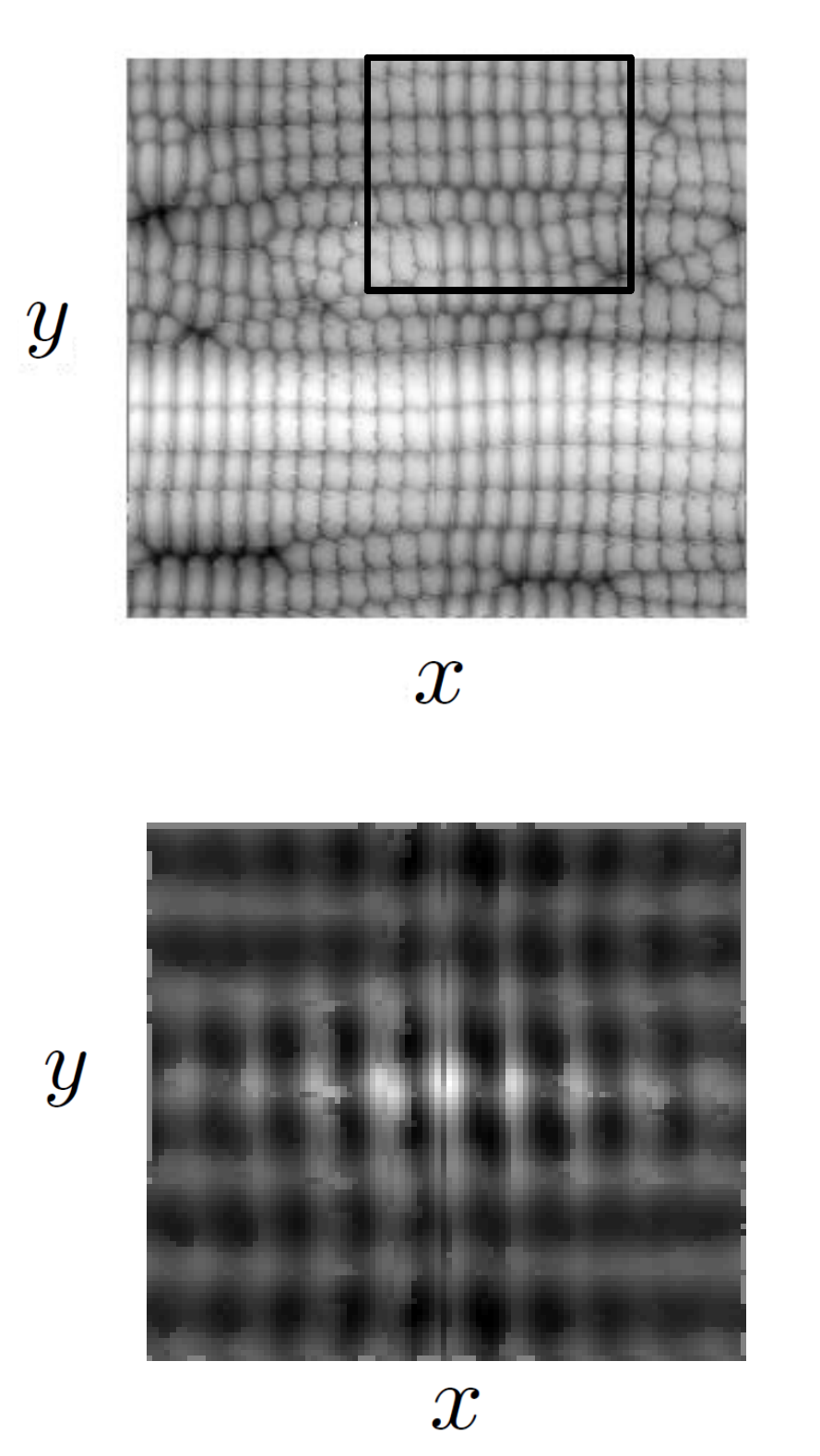}
\end{center}
\end{minipage}\hspace*{ 0.01\linewidth}
\begin{minipage}[c]{0.333\linewidth}
\begin{center}
{ $\beta_x=0.5$}
\includegraphics[width=\linewidth]{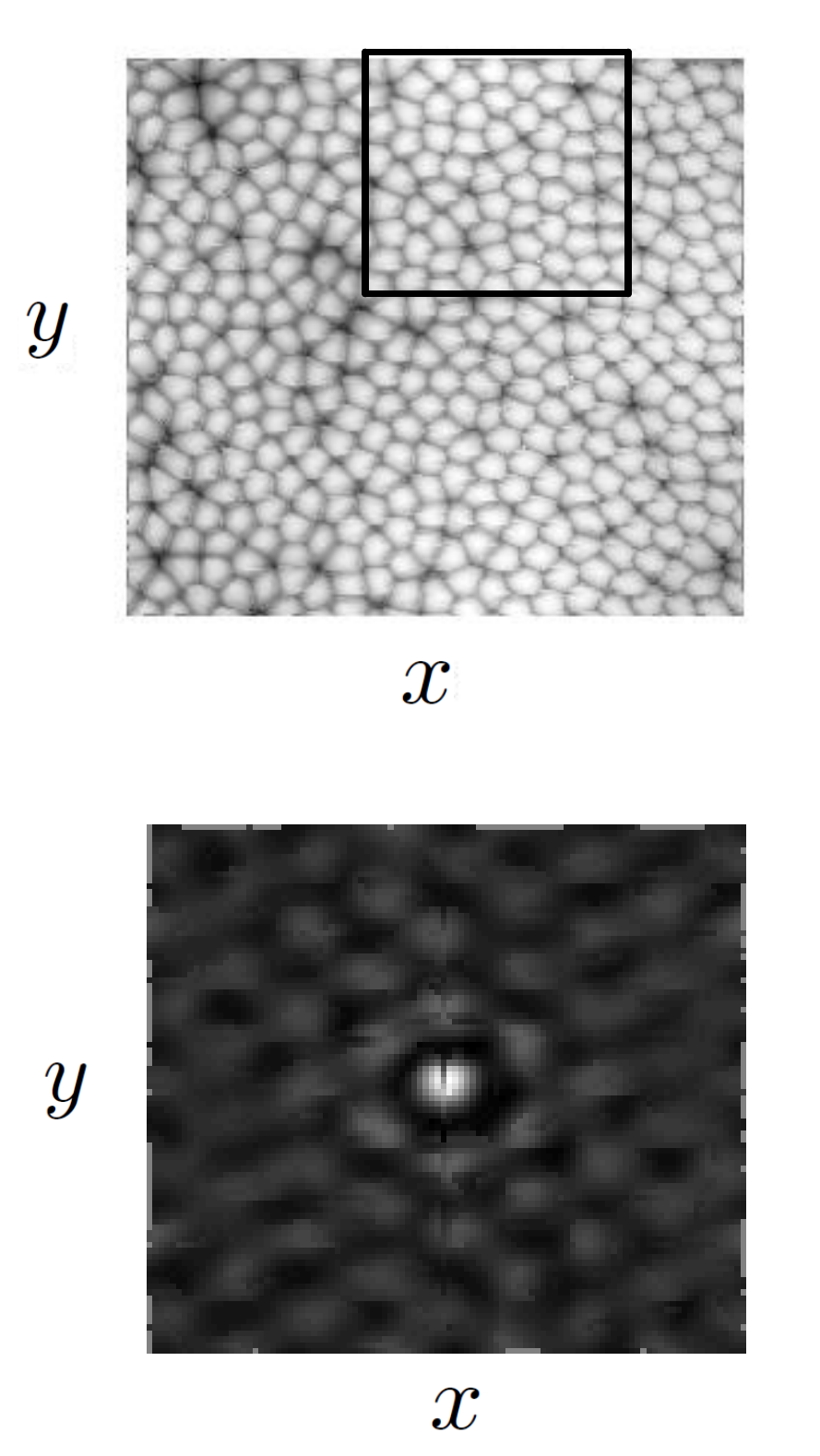}
\end{center}
\end{minipage}\vspace*{ 0.01\linewidth}
\end{center}
\caption{Top-view surface morphologies (top row) predicted by Eq.~(\ref{eq:ec_normal_anis4}) at $t=1000$ for $\alpha_x=0.5$, $r_0=5$, and different values of $\beta_x$ (see legends). Corresponding normalized autocorrelation functions (bottom row) computed over the indicated squares of size $100 \times 100$.}
\label{fig:AutoR_transition}
\end{figure}

Enhancement of local square order can be achieved bringing together the previous property with the fact that local order is improved for relatively small $r_0$ values. Thus, Fig.\ \ref{fig:AutoR_transition} displays the surface morphologies obtained for $r_0=5$ and different values of $\beta_x$ and their corresponding autocorrelation maps. The symmetry of the short-range order of the pattern can be identified easily in the autocorrelation map, which has been calculated for the ($100 \times 100$) black boxes indicated. Indeed, since the morphology is disordered at long distances, some of the local order information is lost when the height autocorrelation function is computed in the whole domain. At any rate, Fig.\ \ref{fig:AutoR_transition} shows how dots with square-symmetric short-range order can actually occur for intermediate values of $\beta_x$ when the elongation along the $y$ direction exists but is not excessively pronounced. 


Closer inspection of Fig.\ \ref{fig:AutoR_transition} actually suggests that up to three main types of patterns can be expected for Eq.\ \eqref{eq:ec_normal_anis4}, depending on the value of $\beta_x$: Ripples (with a dotted substructure) and dots with square or with hexagonal short-range order. Moreover, as we have already seen, the degree of local order of the pattern can be enhanced by tuning the value of $r_0$. 
Indeed, the three main different patterns just mentioned can be clearly distinguished in Fig.\ \ref{fig:AutoR_transition}, where $r_0=5$ has been fixed.
In the case of isotropic surface diffusion ($\beta_x=0.5$), dots with $\ell_x=\ell_y$ group into hexagonal short-range order, where each dot tends to be in the center of a hexagon formed by the nearest neighbor dots, and local regions tend to have the same average height. For intermediate values $\beta_x \in [0.25,\, 0.3]$, square-ordered elongated dots with $\ell_y>\ell_x$ occur. Moreover, for these parameter values the surface heights becomes more heterogeneous, different local regions presenting different average heights. For even lower values of $\beta_x$, a ripple structure appears, with a periodicity along the $y$-direction. Again this morphology displays quite heterogeneous average heights in different regions, while it still features a short-scale structure of rather elongated dots which are quite ordered along the $x$-direction. Hence, decreasing the value of $\beta_x$ induces a transition from short-range hexagonal, to square and then to rectangular ordering of the dots.

An analogous transition between hexagonal and square patterns has been studied in Ref.~\onlinecite{Gollwitzer2006} for the case of magnetic fluids under applied magnetic fields. In this work the authors employ an angular correlation function that makes use of the discrete Fourier transform of the height field in order to characterize the (hexagonal or square) symmetry of the pattern, and thus assess morphological transitions under changes in external parameters. Here, we define a similar angular correlation function, but relative to the values of $R_N$, rather than those of $h(\boldsymbol{x}, t)$. Specifically, the angular autocorrelation function, $P (\psi, t)$, which we propose to quantify the pattern order is
\begin{equation}
	P (\psi, t) =  \frac{1}{L_w^2}\int [R_N (\boldsymbol{x}, t) R_N (\boldsymbol{M}(\psi) \boldsymbol{x}, t)  - \bar{R}_N^2 (t)] d \boldsymbol{x},
\end{equation}
where $\bar{R}_N (t)$ is the space average of the autocorrelation function, $R_N(\boldsymbol{x}, t)$, over a square window of lateral size $L_w$, and as above $\boldsymbol{M}(\psi)$  is the counterclockwise rotation matrix of angle $\psi$. The function $P (\psi, t)$ thus measures the height correlation at every position in the considered domain and compares it with the result obtained at a position which is rotated by an angle $\psi$. We further define the normalized angular autocorrelation function as $P_N (\psi, t)=P (\psi, t)/P (0^\circ, t)$. The reason for considering an area of lateral size $L_w<L$ is because some rotated points $\boldsymbol{M}(\psi) \boldsymbol{x}$ could remain out of the considered domain for a square grid. Besides, due to the global disorder of the patterns induced by the KPZ non-linearity, the short-range order of the pattern needs to be quantified in smaller areas. 

\begin{figure}[!t]
\begin{center}
\includegraphics[width=9cm]{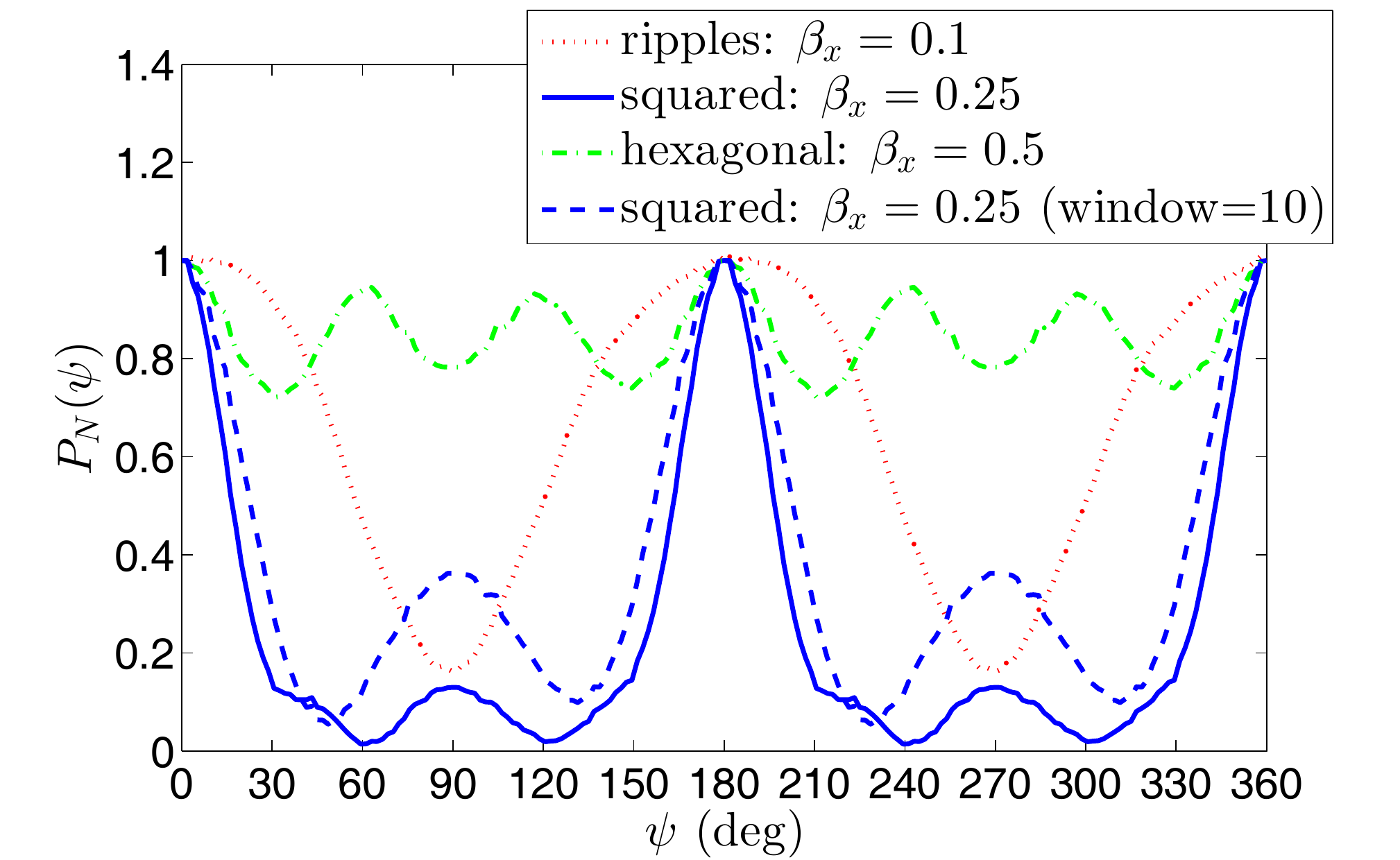}
\end{center}
\caption{(Color online) Normalized angular autocorrelation function, $P_N (\psi,t)$, at $t=1000$ with $L_w=16$ for the morphologies in Fig.\ \ref{fig:AutoR_transition} in which $\beta_x=0.1$ (dotted red line), $\beta_x=0.25$ (solid blue line), and $\beta_x=0.5$ (dash-dotted green line). The dashed blue line shows $P_N (\psi,t)$ for $\beta_x=0.25$ with $L_w=10$.}
\label{fig:Auto_angle}
\end{figure}

The normalized angular autocorrelation functions corresponding to the three basic morphologies shown in Fig.~\ref{fig:AutoR_transition} are displayed in Fig.~\ref{fig:Auto_angle}.
The local extrema in $P(\psi,t)$ signal how well correlated are the points in the morphology with those rotated by an angle $\psi$. In all cases, due to the system symmetry under a 2D space inversion $(x,y)\rightarrow (-x,-y)$, this function is periodic with $180^\circ$ period. For $\beta_x=0.1$ (dotted red line) ripples form and the only rotation that leaves the system unchanged is precisely one with $\psi=180^\circ$, hence the maxima in $P(\psi,t)$ as a function of $\psi$ are separated by this value. A different behavior is found for $\beta_x=0.25$ (solid blue line), when a dot pattern with square-symmetric order appears. In this case the distance between consecutive maxima of $P(\psi,t)$ is $90^\circ$, since the height correlation is itself maximized after a rotation of $90^\circ$ for a pattern with this type of order. Note that, due to the large-scale disorder of the morphology, secondary maxima are smaller than one, indicating a smaller degree of correlation. Improved correlation values for the secondary maxima are obtained for $\beta_x=0.5$ (dash-dotted green line), when dots group into short-range hexagonal order. In this case the angular autocorrelation function suggests the best correlation among surface points for $\psi=60^\circ$ and $\psi=120^\circ$, signaling hexagonal symmetry in the dot arrangement.

We should remark that the size of the patterns and the spatial range of the order can be different in each parameter regime, which requires suitable window sizes for appropriate assessment. In particular, the square-order dots pattern reaches smaller distances than the ripples or the hexagonal dots pattern, so that the maxima of the corresponding angular autocorrelation function are smaller if the same lateral window size is employed. As an example, the angular autocorrelation function for $\beta_x=0.25$ using a smaller window is shown by the dashed blue line in Fig.\ \ref{fig:Auto_angle}. We note that this window size is more appropriate to measure the square-order pattern, since the peak values of the angular autocorrelation at $\psi=90^{\circ}$ and $\psi=270^{\circ}$ are larger and, therefore, the square pattern can be identified better.

\section{Discussion and Conclusions}
\label{Concl}

%
%
%


Motivated by the experimental results in Refs.~\onlinecite{Kim2011} and \onlinecite{Kim2013}, we have generalized the eKS model proposed in Refs.~\onlinecite{Castro2005} and \onlinecite{Munoz-Garcia2006} for anisotropic materials considering anisotropic surface diffusion. We have obtained a continuum equation for the surface height, Eq.\ \eqref{eq:ec_normal_anis2}, which, in contrast to the isotropic eKS model, presents anisotropies caused by the heterogeneous diffusivities along each substrate direction. This model allows to reproduce rippled and square-ordered patterns for normal incidence, akin to those observed in IBS of metals.\cite{Rusponi1998a,Valbusa2002,Kim2011,Kim2013} In particular, in the experiments of Refs.~\onlinecite{Kim2011} and \onlinecite{Kim2013}, prepatterned gold targets were bombarded at normal incidence with Ar$^+$ ions. Although the initial ripples influence the pattern formation substantially  ---to the extent that dots align preferentially along preexisting ripple ridges leading to so-called nanobead structures, (semi)quantitatively described by the eKS equation---, rows of such nanobeads tend to further align in such a way that, on average, each bead has four nearest-neighbor beads: Two along the same row and two in the adjacent bead rows. Such short-range ordering had remained beyond description by the eKS model, while it is similar to what is obtained in Fig.\ \ref{fig:AutoR_transition}, compare e.g.\ with Fig.\ 1(c) in Ref.\ \onlinecite{Kim2013}.

Systematic numerical integration of the anisotropic model at normal incidence was carried out, which has provided indications on the effect of the various terms in the equation and, consequently, of the underlying mechanisms behind each one. In particular, we have focused on observables such as the roughness $W$, the wavelengths along the two independent directions, $\ell_x$ and $\ell_y$, and the type of pattern and order that ensues. Two main characteristics should be highlighted: {\em (i)} Eq.\ \eqref{eq:ec_normal_anis2} is able to predict patterns with $\ell_x \neq \ell_y $ and {\em (ii)} this continuum model can also predict patterns with short-range square order.
Both features have been observed in IBS of metals under normal incidence \cite{Rusponi1998a,Valbusa2002,Kim2011,Kim2013} and had not predicted by previous models. We have also introduced and angular correlation function which has been proven to usefully characterize quantitatively the pattern symmetry.
Furthermore, since the parameters of the equation depend explicitly on physical conditions, it could be possible to design specific experiments to control the resulting pattern if different geometrical properties are required for applications. It is worth mentioning that square patterns have also been observed in IBS experiments on semiconductors (Si and Ge) when metallic contaminants are co-deposited.\cite{Frost2004,Frost2008surface} In those cases, the role of metals is two-fold: on the one hand they trigger pattern formation even for angles below a critical one (a feature that is not observed on {\em clean experiments}\cite{Castro2012a}) and, on the other hand, they introduce anisotropy and, as predicted by our model in the present work.

On general grounds, these conclusions seem to substantiate further the applicability of two-field models like Eqs.\ \eqref{eq:couR}-\eqref{eq:Gamma_ad} for IBS of metallic systems in the erosive regime. One relevant question in this connection is whether (anisotropy-enhanced) short-range order of the type predicted by this model suffices to account for all of the experimental morphologies, or else if stronger ordering properties are required, akin to those found e.g.\ in IBS of binary materials.\cite{Pearson2015}


Finally, the type of model and derivation that we have employed may actually be helpful in two additional contexts in which surface anisotropies play a role. One is IBS of metallic systems under diffusive conditions \cite{Valbusa2002} or of semiconductor targets at high temperatures.\cite{Ou2013,Ou2014,Ou2015} In both cases the (anisotropic) crystalline structure proves to be of paramount importance. Model \eqref{eq:couR}-\eqref{eq:Gamma_ad} should probably be generalized in order to account for anisotropic surface tension and diffusion, allowing for non-linear contributions. The second context is that of surface nanopatterning by ion implantation, in which anisotropic surface diffusion terms have been invoked in order to account for experimental patterns with novel symmetries.\cite{Mollick2014}

\begin{acknowledgments}
This work has been funded through MINECO (Spain) grants FIS2012-38866-C05-01, FIS2012-32349,
and FIS2013-47949-C2-2.
\end{acknowledgments}



\bigskip

\appendix

%

\section{Parameter values of the anisotropic effective equation} \label{app.B}
Following a multiple scales approach which is similar to that employed in Refs.~\onlinecite{Munoz-Garcia2006} and \onlinecite{Munoz-Garcia2008}, the coefficients entering Eq.\ \eqref{eq:eKS_anis} depend on those characterizing the excavation and addition rates $\Gamma_{ex}$ and $\Gamma_{ad}$, being specifically given by\cite{Renedo2013}
\begin{eqnarray*}\label{eq:parameters_eKSanis}
 & & \gamma_x = -\phi \alpha_0 \alpha_{1x}, \nonumber \\ 
 & & \nu_x = \phi \alpha_0 \alpha_{2x} - \frac{\alpha_0^2}{\gamma_0} \bar{\phi} \phi \alpha_{1x}^2, \nonumber \\ 
 & & \nu_y = \phi \alpha_0 \alpha_{2y}, \nonumber \\ 
 & & \Omega_{ij} = \alpha_0 \left( \frac{\bar{\phi} D_{ij}}{\gamma_0} - \phi  R_{eq} \gamma_{2i} \delta_{ij} \right) \alpha_{1x}, \nonumber \\ 
 & & \mathcal{K}_{ijk} = D_{ij} R_{eq} \gamma_{2k} + \alpha_0 \left( \phi R_{eq} \gamma_{2i} \delta_{ij} - \frac{\bar{\phi} D_{ij}}{\gamma_0} \right) \alpha_{2k}, \nonumber \\ 
 & & \lambda_i^{(1)} = -\alpha_0 \phi \alpha_{3i}, \\ \nonumber
 & & \lambda_{ijk}^{(2)} = \alpha_0 \left( \phi R_{eq} \gamma_{2i} \delta_{ij} - \frac{\bar{\phi} D_{ij}}{\gamma_0} \right) \alpha_{2k},
\end{eqnarray*}
where $\delta_{ij}$ is the Kronecker delta, and $i, j, k = x,y$.


\end{document}